\documentclass[review]{elsarticle}

\usepackage{lineno,hyperref}
\usepackage{graphicx,epsfig}
\usepackage{times}
\usepackage{graphics,dcolumn,bm,float}
\usepackage{amssymb,amsmath,rotate,color}
\usepackage[title,titletoc,toc]{appendix}
\DeclareMathOperator*{\dprime}{\prime \prime}
\usepackage{breqn}
\usepackage{bbold}
\usepackage{pgfplots}
\usepackage{float}
\usepackage{tkz-euclide}
\usepackage{braket}
\usepackage{physics}
\usepackage[caption=false]{subfig}
\usepackage[export]{adjustbox}
\usepackage{tikz}
\usetikzlibrary{through,calc}
\usetikzlibrary{positioning}

\modulolinenumbers[5]

\newcommand{\be}{\begin{equation}}
\newcommand{\ee}{\end{equation}}
\newcommand{\ep}{\epsilon}

\newcommand{\bearr}{\begin{eqnarray}}
\newcommand{\eearr}{\end{eqnarray}}
\newcommand{\nn}{\nonumber}

\newcommand{\eps}{\varepsilon}

\newcommand{\up}{\uparrow}
\newcommand{\down}{\downarrow}

\newcommand{\bsq}{{\boldsymbol{q}}}
\newcommand{\bsk}{{\boldsymbol{k}}}
\newcommand{\pr}{\partial}
\newcommand{\bs}{\boldsymbol}

\journal{Journal of JMMM}

\begin{document}

\begin{frontmatter}

\title{Electromagnetic modes from Stoner enhancement}


\author[mymainaddress]{Z. Jalali-Mola}
\ead{jalali@physics.sharif.edu}

\author[mymainaddress,mysecondaryaddress]{S.A. Jafari*}
\cortext[mycorrespondingauthor]{Corresponding author}
\ead{jafari@physics.sharif.edu}
\ead[url]{http://physics.sharif.edu/~jafari/}

\address[mymainaddress]{Department of Physics, Sharif University of Technology, Tehran 11155-9161, Iran}
\address[mysecondaryaddress]{Center of excellence for Complex Systems and Condensed Matter (CSCM), Sharif University of Technology, Tehran 1458889694, Iran}
 



\begin{abstract}
Systems with substantial spin fluctuations, can internally dress the polarization function by ladder 
diagram of Stoner (spin-flip) excitations. This process can drastically modify the electromagnetic response. 
As a case study we provide detailed analysis
of the corrections to the non-local optical conductivity of both doped and undoped graphene. 
While the resummation of ladder diagram of Stoner excitations does not affect the TE mode in
doped graphene, it allows for a new {\em undampled} TM mode in undoped graphene. This is the sole effect of 
corrections arising from ladder diagrams and is dominated by Stoner excitations along the ladder rung
which goes away by turning off the source of spin-flip interactions. 
\end{abstract}


\end{frontmatter}


\section{Introduction}
Spin fluctuations are considered to be important players in strongly correlated systems
and their associated magnetic properties~\cite{Moriya}. 
These sort of fluctuations are among the possible scenarios for the explanation of Cooper pairing
in strongly correlated systems, including the recent Iron based superconductors~\cite{scalapino}. 
Placing a material with strong spin fluctuations such as a spin ice in proximity to a metallic layer
has been proposed as a mechanism to customize the electron-electron interaction~\cite{Kim-qm}. 
In itinerant systems all one needs is a strong enough source of spin fluctuations. This source
is nothing but the short range interaction called the Hubbard $U$ which prohibits simultaneous presence of 
two electrons at the same orbital by billing a high enough cost for double occupancy as $Un_{i\up}n_{i\down}$. 
Here $i$ refers to an atomic orbital localized in a given site $i$ and
$n=c^\dagger c$ is the fermionic 
occupation number. To see why the short range part of Coulomb interaction can so efficiently generate
spin flip, simply rewrite $Uc^\dagger_{1\up} c^\dagger_{2\down} c_{3\down} c_{4\up}$ 
where $1\ldots 4$ are appropriate momentum indices as $-Uc^\dagger_{1\up} c^\dagger_{2\down}  c_{4\up}c_{3\down}$.
The later form can be interpreted as attraction in the spin-flip or Stoner channel. This is how short-ranged Coulomb interactions
can lead to spin-flip processes and cause spin-flip particle-hole (PH) fluctuations.

\begin{figure}[t]
     \begin{tabular}{cc}
       \subfloat (a){\includegraphics [width = .25\textwidth]{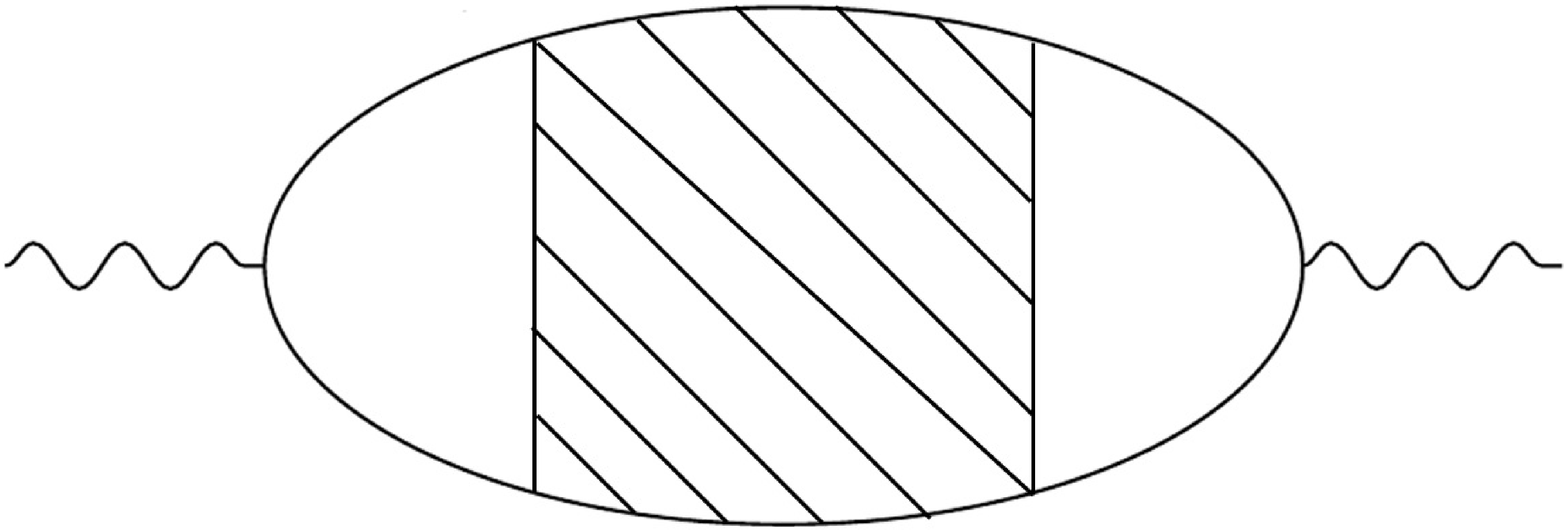}} &
       \subfloat (b){\includegraphics[width = .60\textwidth] {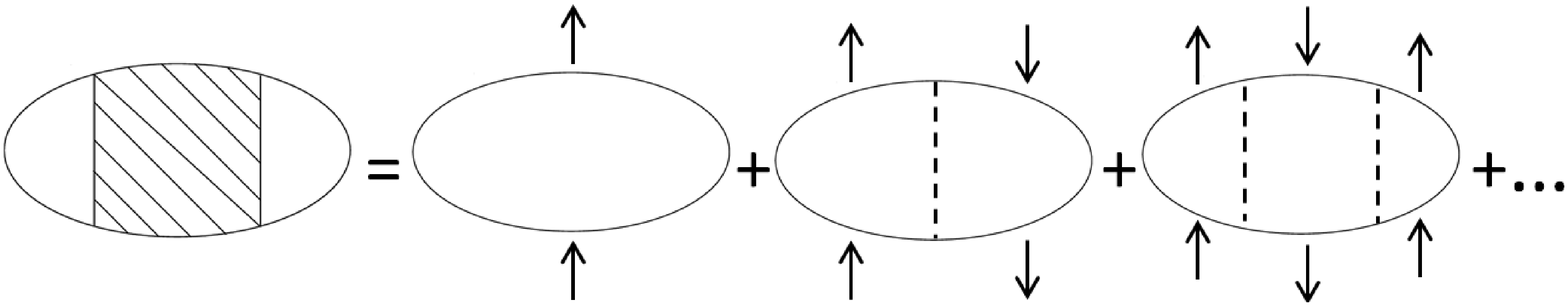}}
       \end{tabular}
      \caption{ (a) General polarization function that determines coupling of any system
      with external photons. (b) Ladder diagrams corresponding to Stoner (spin-flip) excitations
      across the rung. Note that for particle-hole bubbles, particle and hole run in
      opposite directions along the rung. 
      Keeping only the first term in the right hand side corresponds to the 
      commonly used random-phase approximation.
      }
\label{bubbles.fig}
\end{figure}

How do the spin fluctuations of materials manifest in the propagation of electromagnetic (EM) modes?
Put it differently, is there a optical or EM way of directly probing the spin-fluctuations?
The EM response of any system is determined by a fermion bubble to which two external photon propagators are attached 
(Fig.~\ref{bubbles.fig}a). 
The bubble itself can be internally dressed via the so called ladder diagrams (Fig.~\ref{bubbles.fig}b). 
The popular approximation known as random phase approximation amounts to ignoring all the ladder diagrams. 
This is equivalent to keeping only the first (empty bubble) diagram in Fig.~\ref{bubbles.fig}b. 
However, if the materials properties are such that the Hubbard $U$ is strong enough to generate
strong spin-fluctuations of the type described above, it is important to encode them into appropriate
ladder daigram resummation. This is is well-known in the context of high temperature superconductors~\cite{Scalalpinoann}. 
Once a PH pair is created by a light beam, 
the Hubbard $U$ takes care of the spin flips across the rung of the ladder. 
Before the PH pairs recombine to emit back a the photon, 
a resonance enhancement of spin-flip processes known as Stoner enhancement can give rise
to a singularity in the response to the EM radiation. This effect as we will show in great
detail in present work, drastically modifies the propagation of EM radition
in systems that host strong spin fluctuations.
When a Hubbard vertex is inserted on the way of a PH pair running across the rungs of a ladder, 
irrespective of whether the spin is flipped or not, it generates a fluctuation
of the very same electric charge, and hence is expected to quite directly 
affect the dielectric properties. The essential difference between the spin-flip and 
non-spin-filp ladder diagrams will be the sign $\pm U$ of the basic interaction vertex. 

Is there a simple platform that allows for interesting EM properties, and at the same time
hosts strong enough fluctuations of spins?
One of the exciting materials of the past decade has been the two-dimensional graphene
which has attracted great deal of attention during recent years. 
This mono layer material consists of carbon atoms which are arranged in honeycomb lattice and 
the energy spectrum of this material consists of two Dirac cones in the Brillouin zone where 
conduction and valence bands linearly touch each other~\cite{geim,castro,novoselov}. Intrinsic graphene is characterized 
by zero carrier concentration .i.e.,  Fermi surface at zero temperature shrinks to zero (corresponding to $\mu=0$). 
In this case single particle spectrum is characterized with Fermi velocity which is $1/300$ of the velocity of light 
in the vacuum~\cite{vf}.
 In extrinsic graphene, the outstanding role of many body interactions is to renormalize the Fermi velocity of 
graphene~\cite{jafari2009,Anand2015,Anand2016} which is strongly dependent on carrier concentration. For typical values of carrier concentrations  
$n\approx10^{12} cm^{-2}$, the Fermi velocity is same as intrinsic case~\cite{vf,ultralow,vf3}. In the 
ultralow doping regime, down to three orders of magnitude less then the above typical values, 
a logarithmic dependence of Fermi velocity on the carrier concentration can be detected~\cite{ultralow}
which signifies the importance of many-body interactions.

First of all, 
among many unconventional properties of graphene, its EM response is also different from a normal 
2D electron gas~\cite{stauber2010,stauber2013,principi,scholz,hwang,wunsh,stauber2008,gusynin-shaparov2008,koshino}
at least in two respects: (i) Unlike normal 2D electron gas which only admits transverse magnetic (TM) mode, 
the chiral electron gas in graphene allows for the propagation of transverse electric (TE) mode 
in the terahertz (THz) range~\cite{mikhailove,review,jablan,merano} which is not possible in non-chiral 2D electron gas.
(ii) The regularized electromagnetic response of graphene is basically characterized by dimensionless energy 
variable $\hbar\omega/ \ep_F$ and dimensionless momentum variable $q/k_F$, as the only time and
length scales of a Dirac theory at non-zero density are set by the Fermi energy $\ep_F$ and Fermi wave vector $k_F$. 
This means that the smallness of wave-vectors is naturally measured with respect to the $k_F$.
In ultra-low doped graphene $k_F$ is small and, hence, the ratio $q/k_F$,
for typical THz electromagnetic waves, can be large. Thus, one needs
to take into account the dispersion, i.e. the wave-vector dependence, and resulting
anisotropy, of the conductivity tensor, $\sigma(\bsq, \omega).$
Therefore a simple gate voltage provides a handle to explore the non-local (i.e. $\bsq\ne 0$)
aspects of the EM response in this system~\cite{principi,scholz,mikhailove}. 

Secondly, a recent {\em ab-initio} estimates of the short range interactions in graphene suggests remarkably 
large value of $U\sim 10$ eV~\cite{wehling}. This is expected to generate a substantial amount of spin fluctuations.
With such a large Hubbard $U$ the spin fluctuations
become so large that it has been proposed that the ground state of Hubbard model on the 
honeycomb lattice becomes a spin liquid~\cite{meng}. 
The role of spin-flip fluctuations in graphene has been extensively studied by one of us in the past, and the general 
picture is that the the 
cone-like nature of single-particle excitations gives rise to a window below the particle-hole continuum 
which is void of free particle-hole excitations. 
This window provides a chance to develop a coherent pole in the ladder summations 
which can be interpreted as a bound state of particle-hole excitations in the spin-flip channel. 
This may happen in both undoped~\cite{Jafari2002, Jafari2004,Jafari2012,Ebrahimkhas2009,Hedegard} and doped 
graphene~\cite{Jafari}. 
In this work we would like to study the effect of such ladder diagrams in 
the electromagnetic response of graphene, and in particular to focus on the special
role played by the spin-flip channel of particle-hole fluctuations. 

For momentum dependent interactions in a limited range of parameters the dressing
of fermion polarization bubble with non-spin-flip ladder diagrams has been considered by others~\cite{mishchenko}.
It turns out that the ladder corrections (in the non-spin-flip channel) give rise to a new {\em damped} TM mode.
In this work we would like to study a much more manifest form of this effect which unlike the 
previous study~\cite{mikhailove} rests on: (i) short range interactions and (ii) the spin-flip
channel. It turns out that with short range interactions, the dominant effect is due to 
spin-flip processes, and the non-spin-flip ladder corrections will become irrelevant. 
Note that the short range (i.e. momentum independence) of interaction brings in a great
technical simplification: The ladder diagrams can be easily summed into a simple
{\em RPA-looking} expressions that are actually ladder diagrams. 
Otherwise the resummation of ladder diagrams for momentum-dependent vertices is rather involved, 
and can only be performed under sever approximations.

Let us advertise the main result of incorporation of ladder diagrams into the polarization bubble in graphene:
The first and straightforward message will be that the Hubbard interaction being longitudinal (density-density)
interaction does not give any corrections to the TE mode. This becomes transparent when 
we represent the conductivity tensor in terms of its longitudinal and transverse components. 
Therefore the spin fluctuations do not affect the TE mode~\cite{mikhailove} of doped graphene. 
For the TM mode,
we find that although in doped graphene, the spin-flip or Stoner particle-hole excitations do not find a chance 
to develop a coherent pole at small momenta~\cite{Jafari}, 
these fluctuations are still able to modify the TM mode by taking advantage of the effective minus sign
generated in re-arranging the interaction into spin-flip form.
The general effect of this minus sign is to
reduce the energy of the TM mode at any given wave vector $\bsq$. The result of such a reduction in the
energy of the TM mode becomes spectacular in the undoped graphene: For a chiral electron gas if we
ignore the corrections due to ladder diagrams of Fig.~\ref{bubbles.fig}, the Maxwell equations give no undamped solutions 
for the TM mode as the mode energy strongly overlaps with the continuum of free particle-hole excitations, 
meaning that the longitudinal density oscillations of the TM mode decay and emit free particle-hole pairs. 
However, dressing the empty bubble by ladder of Stoner processes
drastically changes this picture, and brings the energy of the TM mode below the particle-hole continuum. 
Therefore the resulting TM mode will be protected from Landau damping. In this way the
Stoner particle-hole fluctuations serve as a unique mechanism to generate a branch of TM mode in 
{\em undoped} graphene which would have been impossible if the electrons had no spin to flip.
This spectacular effect can be considered as optical proble of the spin fluctuations. 

 This paper is organized as follows. In Sec. II we start with the graphene band structure, and formulate the
current-current response in its tensor form and represent it in terms of two independent components, namely
longitudinal and transverse ones. 
In section III starting from Maxwell's equations we derive the dispersion equations for TE and TM modes
from which it will be manifest that the TE mode does not receive corrections from ladder resummation,
while the TM mode can be modified by resummation of Stoner ladder diagrams arising from short range Hubbard interactions. 
Building on equations of section III, in
section IV we first revisit the problem of TE and TM modes in non-interacting graphene. In section V we 
turn on the Hubbard interaction and use the ladder diagram resummation to correct the equation of TM mode. 
We end in section VI with a summary and discussion.

\section{\label{sec:level1}Current-current correlation tensor for noninteracting 2D Dirac model}
Linear dispersion of non interacting graphene (near Dirac points $\boldsymbol{K}$ or $\boldsymbol{K'}$)  
is described by the following Hamiltonian in the creation and annihilation operator representation,
\be
\label{hamiltonian}
\hat{H_0}(\bsk)=\hbar v_F \sum_{s}\hat{\psi}^\dag_{s,\bs{k}}  (\bs{\sigma}. \bs{k}) \hat{\psi}_{s,\bs{k}}.
\ee
Here $\hat{\psi}^\dag_{s,\bs{k}} =( \hat{a}^\dag_{s,\bs{k}} \hat{b}^\dag_{s,\bs{k}})$
is the spinor consisting of creation operator for an electron at momentum $\bs{k}$ and spin $s$ in either
of the sublattices A or B, $\bs{\sigma}=(\sigma_x , \sigma_y)$ denotes Pauli matrices in the space of two sublattices, 
$\bs{k}=(k_x, k_y)$, $k=\sqrt{k_x^2+k_y^2}$ and Fermi velocity $v_F\simeq \frac{c}{300}$ with $c$ being the speed of light~\cite{castro}. 
At ultralow doping the velocity can be enhanced by interaction effects~\cite{ultralow}.
Since the Dirac nodes around which the linearized dispersion holds corresponds to a non-zero momentum in the
Brillouin zone, by time reversal symmetry, there should be another Dirac valley at opposite momentum.
Therefore the complete set of low-energy degrees of freedom consists in additional valley degeneracy. 
If we label the two valleys with $\tau=\pm1$ the dispersion around the two valleys will be given by the $2\times 2$
matrix $\sigma_x k_x+\tau\sigma_y k_y$ which give rise to identical dispersion relation. Switching between
the valleys amounts to $k_y\to -k_y$ transformation. This transformation does not affect the propagation 
of electromagnetic modes in graphene. Therefore we consider only one valley, and as long as the propagation
of electromagnetic modes in graphene is concerned, the presence of other valley
can be taken into account by a multiplicative factor of $2$. 
As far as non-interacting electrons are concerned, the similar argument applies to spin degeneracy. 
However when the particle-hole fluctuations are included, 
since particles and holes are spin-half fermions, with respect to the  spin of the particle-hole
pair there are two channels for the fluctuations of particle-hole pairs, namely Stoner (spin-flip) and
non-spin-flip. 
As argued in the introduction, the sign of interaction in these two channels are different,
and therefore upon resummation of the series of ladder diagrams in both channels, the two channels 
split off, and the role of spin is not a factor of $2$ anymore.
Therefore
the multiplicative factor of $2$ for non-interacting electrons should be dropped when dealing with the
separate contribution of the above two channels in presence of strong Hubbard $U$ term.

 The eigenvalue equation for a single valley Hamiltonian is given by,
 \be
 \label{eigenvalue}
 \hat{H}_0(\bsk)\ket{\bs{k},\pm}=\pm \hbar v_F k \ket{\bs{k},\pm},
 \ee
 where,
 \be
 \nonumber
 \ket{\bs{k},\pm}= \begin{pmatrix} 1 \\ \pm e^{i\phi_k}\end{pmatrix}.
 \ee
  Here positive and negative eigenvalues correspond to valence and conduction bands respectively, which touch each other  in Dirac point and $\phi_k$ is the polar angle of $\bs{k}$ with respect to the the $k_x$ axis. In the case of doped graphene when we measure energies with respect
  to the Fermi level, the energy eigenvalues will be given by $E_{\pm,d}(k)=\pm \hbar v_F k-\mu$ where $\mu=\hbar v_F k_F$ is
the chemical potential, the subscript $d$ stands for doping, and we have implicitly assumed electron doping (i.e., $\mu>0$). 
  The current operator corresponding to this Hamiltonian is,
  \be
     \hat{\bs{j}}_\bsq=v_F \sum_{\bsk} \hat{\psi}^\dag_{s,\bs{k}}  \bs{\sigma} \hat{\psi}_{s,\bs{k+q}}.
  \ee
One could augment the above relation to include the zeroth ("time") component such that 
${\hat{j}_\bsq^0}$ proportional to $\sigma_0$ (the unit matrix) which after dividing by the Fermi velocity 
gives the density operator. 
 The electromagnetic response of the system is given by the equilibrium correlation function which in the linear response 
 theory is expressed by the Kubo formula,
  \be
     \Pi^{\alpha\beta}(\bs{q},\omega)=\int\frac{dt}{2\pi i}e^{i\omega t} \theta(t)\langle [\hat{\bs{j}}_{\bs q}^\alpha(t), \hat{\bs {j}}_{-\bs q}^\beta(0)] \rangle,
  \ee
  where $\theta(t)$ is the step function. First of all, the above current-current correlation
  function is a tensor quantity. Secondly this relation holds in presence of interactions as well. 
  Let us elaborate on how the combination of these two properties can affect the electromagnetic
  response of graphene: Given a vector quantity ${\bs q}$ characterized e.g., by the 
  polar coordinates $(q,\theta)$, the most general form of the Cartesian components of a rank two symmetric
  tensor quantity is given by,
\begin{align}
   &T^{xx}=c+a\cos^2\theta + b \sin^2\theta\sim c+aq_x^2+bq_y^2,\nn\\
   &T^{yy}=c+a\sin^2\theta + b\cos^2\theta\sim c+aq_y^2+bq_x^2,\nn\\
   &T^{xy}=(a-b) \sin\theta\cos\theta\sim (a-b) q_x q_y,\nn
\end{align}
where $a,b$ are scalar quantities with respect to the rotation of the coordinates.
It turns out that $a$ ($b$) is the longitudinal (transverse) component of the tensor. 
Note that on pure mathematical grounds, in systems with broken inversion symmetry terms of the $\lambda_{\alpha\beta\gamma}q^\gamma$
are also possible which are related to natural optical activity and give rise to
chiral effects~\cite{Pesin2015}. But in the case of grahpene such terms are absent. 
Propagation of TE and TM modes are given by two different functions as in Eq.~\eqref{te-simp} and ~\eqref{tm-simp}. 
Ignoring the tensor character by setting $\bs q=0$~\cite{mikhailove} amounts to assuming an isotropic
form for the tensor (i.e. assuming it is proportional to unit matrix). This in turn will erroneously give the
same functional form in the dispersion relation of both electromagnetic modes. 
Therefore it is necessary to consider the $\bsq$ dependence of the conductivity tensor.
Once this is done, we obtain Eq.~\eqref{te-simp} and Eq.~\eqref{tm-simp} which are valid
for both interacting and non-interacting current-current response tensors.
In terms of the above representation of the conductivity tensor the interactions will provide
corrections to the longitudinal ($a$) coefficient, leaving the transverse channel ($b$) intact. 
Only interactions of the form in the Thirring model are
able to provide interaction corrections in both channels. 

Let us now write down the current response and see if it is of the above general form or not.
We start by calculation of the tensor components for the two-current correlation function~\cite{principi,scholz}
which is defined by the following Lehman representation,
\bearr
\Pi^{ij}(\boldsymbol{q},\omega)&&=  g  \lim\limits_{\eta\rightarrow 0}  \sum_{k,\lambda , \lambda'=\pm} \frac{n_{k,\lambda}-n_{k+q,\lambda'}}{\hbar \omega+E_{k,\lambda}-E_{k+q,\lambda'}+i\eta}\nn\\&& \times\bra{\bs{k},\lambda}\bs{\sigma}^i\ket{\bs{k+q},\lambda'} \bra{\bs{k+q},\lambda'}\bs{\sigma}^j\ket{\bs{k},\lambda}.
\eearr
Here, $g=\frac{ g_s g_v v_F^2}{A} $ which  $g_s$ and $g_v$ respectively introduce spin and valley degeneracy each being equal to $2$ in the case of Dirac fermions in graphene and $\eta$ is an infinitesimal and positive quantity and $i,j$ can take on $x,y$ directions and $A$ is the area of the sample. Fermi distribution function is labeled by $n_{k,\lambda}$ which is a step function at  zero temperature, $E_{k,\lambda}$ is the linear dispersion of graphene, $i$ and $j$ are the component of current operator which are given in terms of Pauli matrices. A derivation of current response function  for doped graphene including mass gap has been done by Scholz \textit{et al.}~\cite{scholz}. We provide our own derivation with emphasize on tensor character. 

As a result of expanding overlap of Pauli matrices between eigenvectors, current response function reduces to:
\bearr
&&\Pi^{ij}(\boldsymbol{q},\omega)=
\nn\\&& g \lim\limits_{\eta\rightarrow 0}  \sum_{k,\lambda , \lambda'=\pm} \frac{n_{k,\lambda}-n_{k+q,\lambda'}}{\hbar \omega+E_{k,\lambda}-E_{k+q,\lambda'}+i\eta}  f^{ij}_{\lambda,\lambda'} \left(\boldsymbol{k},\boldsymbol{q}\right).\label{pikubo.eqn}
\eearr
In this equation, the form factor $f^{ij}_{\lambda,\lambda'} \left(\boldsymbol{k},\boldsymbol{q}\right)$ will be different depending on which elements of current operator is being considered. If current operator components are same (diagonal), the corresponding form factor has the form $f^{ii}_{\lambda,\lambda'}=\left[1\pm\lambda \lambda' \cos\left(\phi_k+\phi_{k+q}\right)\right]/2$ with $+$ ($-$) 
for $i=x~(y)$ component of current. For off-diagonal cases we have $f^{xy}_{\lambda\lambda'}=f^{yx}_{\lambda,\lambda'}=\left[\lambda \lambda' \sin\left(\phi_k+\phi_{k+q}\right)\right]/2$. Here $\phi_k$ and $\phi_{k+q}$ represent the polar angles of the wave vectors 
$\bs k$ and $\bs k+\bs q$ with respect to the $x$ axis, respectively. The direction of $\bs q$ with respect to 
$x$ axis is determined by $\theta$. The required $\cos$ and $\sin$ functions are given by,
\begin{align}
  \label{cos}
  \cos(\phi_k +\phi_{k+q})=\frac{ k\cos(2\phi_k+\theta)+q\cos(\phi_k+\theta)}{|\bs k+\bs q|},\\
 \label{sin}
  \sin(\phi_k+\phi_{k+q})=\frac{ k\sin(2\phi_k+\theta)+q\sin(\phi_k+\theta)}{|\bs k+\bs q|}.
\end{align}
These representation of form factor helps us to express all the components of the conductivity tensor in terms of a single function
$\Pi^{xx}(q,\theta,\omega)$. Before doing any of the integrations in  Eq.~\eqref{pikubo.eqn} it can be cast into the following 
matrix form (see appendix A),
\begin{equation}
  \Pi(q,\theta,\omega)=\frac{1}{2} \begin{pmatrix}  \pi_0+\pi_3 &  \pi_2\\  \pi_2 & \pi_0-\pi_3 \end{pmatrix},
  \label{matrixform}
  \end{equation}
where,
\bearr
   &&\pi_0=\Pi^{xx}(q,\theta,\omega)+\Pi^{xx}(q,\theta+\pi/2,\omega),\\
   &&\pi_2=\Pi^{xx}(q,\theta-\pi/4,\omega)-\Pi^{xx}(q,\theta+\pi/4,\omega),\\
   &&\pi_3=\Pi^{xx}(q,\theta,\omega)-\Pi^{xx}(q,\theta+\pi/2,\omega).
\eearr
This representation explains that in order to find out the current response tensor it is sufficient to  derive general form of first diagonal element of current tensor i.e., $\Pi^{xx}(\bsq,\omega)$, as we will do it in the following. 
\begin{figure}[t]
\centering{\includegraphics[angle=0,width=.65\textwidth]{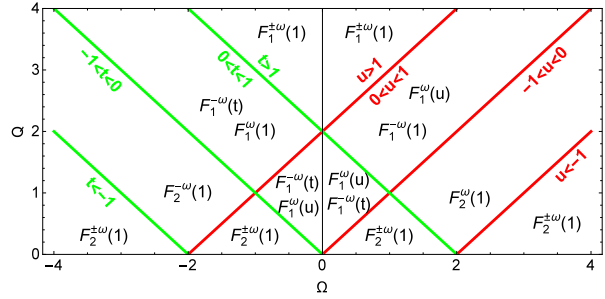}}
\caption{ (Color online) Real part $\Delta \Pi^{xx}(\boldsymbol{q},\omega)$ in different regions. Note that we use $t=(Q +\Omega)/2 $ and $u= (Q-\Omega)/2$ with dimensionless variables $Q =q/k_F$ and $\Omega =\hbar \omega /\ep_F$  where $k_F$ and $\ep_F$ are Fermi wave vector and Fermi energy respectively. The functions $F^{\pm\omega}_1$ and $F^{\pm\omega}_2$ are given in the appendix A.
	}
\label{partitioning.fig}
\end{figure}

The convenient way to do calculation is to subtract current response of undoped graphene defined as $\Pi^{xx}_u(\bsq,\omega)$, from the current response of doped graphene, $\Pi^{xx}_d(\bsq,\omega)$ (the subscripts $u,d$ stand for undoped and doped, respectively). We shall then add it back at the end of calculation.
\be
\label{subtract}
 \Pi^{xx}_d(\boldsymbol{q},\omega)=\Delta \Pi^{xx}(\boldsymbol{q},\omega)+\Pi^{xx}_u(\boldsymbol{q},\omega)
\ee
\bearr
\label{dpiw}
\Delta \Pi^{xx}(\boldsymbol{q},\omega)&&= \Pi^{xx}_d(\boldsymbol{q},\omega) - \Pi^{xx}_u(\boldsymbol{q},\omega)\nn\\&& =  \frac{g_s g_v  v_F^2 }{4\pi^2 \hbar}  \int_{0}^{k_f} k dk d\phi \nn\\&&\times [P(\bs{k},\bs{q},\omega) + P(\bs{k},\bs{q},-\omega)].
\eearr
\begin{figure}[t]
\centering{\includegraphics[angle=0,width=.7\textwidth]{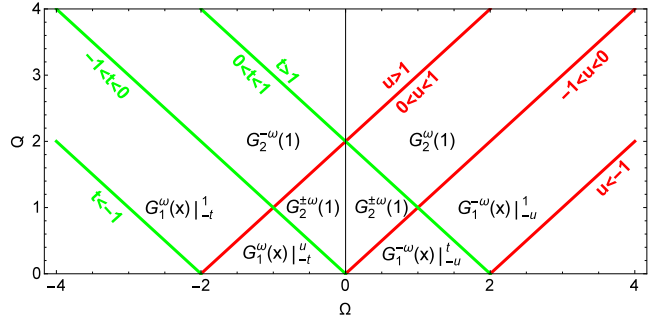}}
\caption{ (Color online) Imaginary part $\Delta \Pi^{xx}(\boldsymbol{q},\omega)$ in different regions. Note that we use $t=(Q +\Omega)/2 $ and $u= (Q-\Omega)/2$ with dimensionless variables $Q =q/k_F$ and $\Omega =\hbar \omega /\ep_F$ where $k_F$ and $\ep_F$ are Fermi wave vector and Fermi energy respectively. The functions $G^{\pm\omega}_1$ and $G^{\pm\omega}_2$ are given in the appendix A.
	}
\label{impartitioning.fig}
\end{figure}
The current response function for the non-interacting undoped graphene can be represented in terms of two functions $F_0$ and $G_0$
as follows,
\bearr
\label{realundoped}
F_0&&\equiv \Re \left[\Pi^{xx}_{0,u}\left(\boldsymbol{q},\omega\right)\right]\nn\\&& =-\frac{g_s g_v }{4\pi \hbar^2 }  \ep_{max}-\frac{g_s g_v}{16 \hbar} C_2 \Theta\left(\left(v_Fq\right)^2-\omega^2\right),\\
\label{imundoped}
G_0 &&\equiv \Im \left[\Pi^{xx}_{0,u}\left(\boldsymbol{q},\omega\right)\right]\nn\\&&= -\frac{g_s g_v}{16 \hbar} C_2 \Theta\left(\omega^2- \left(v_Fq\right)^2\right),
\eearr
where $C_2$ is a function of $(\bs q,\omega)$ defined in the appendix A Eq.~\eqref{c2.eqn} and subscript zero stands for non-interacting graphene. 
Here, cutoff energy of Dirac fermions ($\ep_{max}$) appears by integrating the imaginary part ($G_0$) within a Kramers-Kr\"{o}nig relation. 
When the cut off tends to infinity, this terms leaves an infinity in the response which can be
remedied by normal ordering of the operators employed in calculations of correlators in field theories. A more physical argument
to abandon the cutoff dependent term is as follows:
In order to have gauge invariant response function and due to the diamagnetic sum rule, cutoff term should be ignored because a real physical system can not respond to the longitudinal vector potential in static limit i.e., static longitudinal current response (LCR) function independent of magnitude of $q$ should be zero. In the same way, in the limit of $q\rightarrow 0$ transverse current response (TCR) function equals to longitudinal ones and it should be zero in static limit~\cite{principi,polini,macdonald,vignale}.

The function $\Delta\Pi^{xx}(\bsq,\omega)$ that adds the effect of doping consists in two complex terms: $P(\boldsymbol{k},\boldsymbol{q},\pm\omega)$ (see Eq.~\ref{dpiw}). 
The complete expression and more details of its calculation is given in appendices A and B. We provide the compact form of real (imaginary) part of   $\Delta \Pi^{xx}$ in terms of $F_1^{\pm \omega}(x)$ or $F_2^{\pm \omega} (x)$ ($G_1^{ \pm\omega}(x)$ or $G_2^{\pm \omega} (x)$ ) where $\pm\omega$ 
refer to the first/second term of  Eq.~\eqref{dpiw}, respectively. The regions in the $(q,\omega)$ plane where each function determines the
response is shown in Fig.~\ref{partitioning.fig} and Fig.~\ref{impartitioning.fig} where the dimensionless frequency $\Omega=\hbar \omega/\eps_F$
and wave vector $Q=q/k_F$ are naturally used ($\ep_F$ and $k_F$ are Fermi energy and Fermi wave vector respectively). At the end we need to 
add the result of undoped response function to obtain the final expression for real and imaginary part of current response function. 

Let us emphasize that starting from Eq.~\eqref{pikubo.eqn} and assuming that 
the polar angle of $\bs q$ with respect to the $x$ axis is $\theta$,
the angular dependence of the current response tensor can be separated as,
\begin{equation}
\label{angular res}
  \Pi(\bs q,\omega) = B(q,\omega)  \mathbb{1}+A(q,\omega)  \begin{pmatrix}  \cos2\theta &  \sin2\theta\\  \sin2\theta & -\cos2\theta \end{pmatrix}
\end{equation}
which is the decomposition of a rank 2 Cartesian tensor in terms of its spherical components with 
angular independent complex coefficient $A(q,\omega) $ and $B(q,\omega)$.
 
 \section{\label{sec:level1}electromagnetic response}
Among the interesting features of 2D Dirac materials, their response to electromagnetic fields from the point of 
view of spin fluctuations deserves investigation. 
In normal 2D electron gas only a TM mode can propagate~\cite{stern,falko}, while in the Dirac systems 
already without a ladder resummation correction, the possibility of having a
negative imaginary part for the dynamical conductivity provides a new chance for 
the propagation of TE mode~\cite{mikhailove}. In this section
we would like to explicitly demonstrate the role of non-zero wave vector $\bs q$ in the propagation of electromagnetic
modes in graphene. In the absence of a vector $\bs q$, i.e., when $\bs q=0$, the off-diagonal components of the
conductivity tensor vanish and the diagonal components are equal 
and therefore the only remaining component of the reducible rank two conductivity tensor 
in Eq.~\eqref{angular res} is its scalar part given by 
$B(q=0,\omega)=-i\omega\sigma(\omega)$ where the {\em scalar} $\sigma(\omega)$ is the optical conductivity of
graphene. Therefore the $\bs q=0$ limit misses the entire tensor structure of the conductivity by reducing it to
scalar part. 

Starting from Maxwell's equations, the dispersion relation for the electromagnetic modes in a
two dimensional medium for transverse electric and magnetic modes 
is given by following expressions (for details see Appendix C),
\begin{align}
   &1-\frac{2\pi i \omega}{c^2 \sqrt{q^2-\omega^2/c^2}} \varGamma(\bsq,\omega) =0,~~~~~\mbox{(TE)} \\
   &1+\frac{2\pi i \sqrt{q^2-\omega^2/c^2}}{\omega} \varGamma'(\bsq,\omega)=0,~~~~~\mbox{(TM)}
\end{align}
where the functions $\Gamma$ and $\Gamma'$ are given by,
\begin{align}
   \varGamma(\bsq,\omega)&= \sigma^{yy}(\bsq,\omega) \cos^2\theta+ \sigma^{xx}(\bsq,\omega) \sin^2\theta\nn\\ 
   &-\sigma^{xy}(\bsq,\omega) \sin2\theta\label{gama},\\
   \varGamma'(\bsq,\omega)&=\frac{ \sigma^{xx}(\bsq,\omega)\sigma^{yy}(\bsq,\omega)-\sigma^{xy}(\bsq,\omega)\sigma^{xy}(\bsq,\omega)}{\varGamma(\bsq,\omega)},
   \label{gamap}
\end{align}
which reflects how all the components of the two-current correlation tensor determine the 
propagation of electromagnetic modes in two-dimensional graphene. In the special case of $\bs q=0$, 
both $\varGamma(\bs q=0,\omega)$ and $\varGamma'(\bs q=0,\omega)$ reduce to the scalar conductivity $\sigma(\omega)$
and hence the above dispersion relations for TE and TM modes reduce to those originally 
used in graphene by Mikhailov and Ziegler~\cite{mikhailove}. 
     
  Above relations show angular dependence of $\varGamma$ and $\varGamma'$. Let use eliminate conductivity 
in favor of two-current correlation $\Pi$ by inserting  
$ -i \omega\sigma^{\alpha\beta}(\boldsymbol{q},\omega)=e^2\Pi^{\alpha\beta}(\boldsymbol{q},\omega)$ 
and using the general representation of the conductivity tensor in Eq.~\eqref{angular res} to further
simplify the dispersion relations of TE and TM modes. This gives a very simple and appealing result:
\begin{align}
   \label{te-simp}
   &1+\frac{2\pi e^2 }{c^2 \sqrt{q^2-\omega^2/c^2}} \Pi^{xx}(q\hat{\boldsymbol{y}},\omega)=0,~~~~~\mbox{(TE)}\\
   \label{tm-simp}
   &1-\frac{2\pi e^2 \sqrt{q^2-\omega^2/c^2}}{\omega^2} \Pi^{xx}(q\hat{\boldsymbol{x}},\omega)=0,~~~~~\mbox{(TM)}.
\end{align}
The above simplified version emphasizes that we only need one component $\Pi^{xx}$ of the
current-current correlation tensor to determine the propagation of EM modes in graphene. 
The interpretation of the above equations is that TE and TM modes are governed by 
transverse and longitudinal parts of the two-current response tensor, respectively.
Furthermore in the limit of $\bs q=0$ given that $\Pi^{xx}(\omega)=\Pi^{yy}(\omega)= -i \omega  \sigma(\omega)/e^2$, 
our equations reduce to those used by Mikhailov and Ziegler~\cite{mikhailove}. 

In general an electromagnetic radiation with energies on the scale of eV correspond to wave-lengths
which are typically $10^3$ times larger that atomic distances. In the reciprocal space this would 
imply that the wave vector of such EM modes are a tiny fraction $10^{-3}$ of the Brillouin zone. 
Therefore in terms of the absolute magnitude of the wave vectors of the EM modes it may appear 
safe to approximate $\varGamma(\bs q,\omega)$ and $\varGamma'(\bs q,\omega)$ with their
$\bs q\approx 0$ values which then gives a dispersion relation for the TE and TM modes in terms of 
the uniform optical conductivity $\sigma(\omega)$~\cite{mikhailove}. However the form of 
Dirac equation means that various correlation are only functions of dimensionless $Q=q/k_F$ and $\Omega=\hbar\omega/\eps_F$.
Therefore as far as the response of 2D Dirac system is concerned, whether we can set $\bs q\approx 0$
or not, depends on the relative magnitude of $q$ and $k_F$ and not the absolute magnitude of $q$ which is
very small in Brillouin zone scales any way. In the light of recent developments of ultralow 
doping of graphene, very small values of $k_F$ are attainable~\cite{ultralow}, which then provides access to
finite $\bsq$ electromagnetic response of graphene. Therefore it is timely to
explore aspects of the $q$ (or more precisely $Q$) dependence of the EM response of graphene. 

In the ultralow doping regime, not only the $q$-dependence of the EM response is important, but 
also the many-body interactions become increasingly more important. As a result of stronger
many-body interactions in this regime, among other things the many-body effects enhance the 
Fermi velocity $v_F$ of the charge carriers in graphene~\cite{ultralow}. In the following sections we will
investigate the propagation of TE and TM modes both with and without ladder diagrams due to Hubbard interaction. 
This will be done for both doped and undoped graphene.

\section{Propagation of electromagnetic modes in graphene without ladder corrections}
  In order to study the propagation of TE mode, first of all we need  transverse current response function. 
However, current response is complex function. Therefore when the imaginary part of it is zero, propagation of TE mode 
based on Eq.~\eqref{te-simp} will be governed by real part of the transverse current response. On the contrary when 
imaginary part is nonzero, the mode will be damped due to dissipation of energy to particle-hole pairs.
In this section we will consider graphene without ladder diagram and will focus on the
propagation of EM modes in the absence of Stoner excitations. The role of ladder diagrams will be
considered in the next section.
      \begin{figure}[t]
     \centerline{ \includegraphics[width = .50\textwidth] {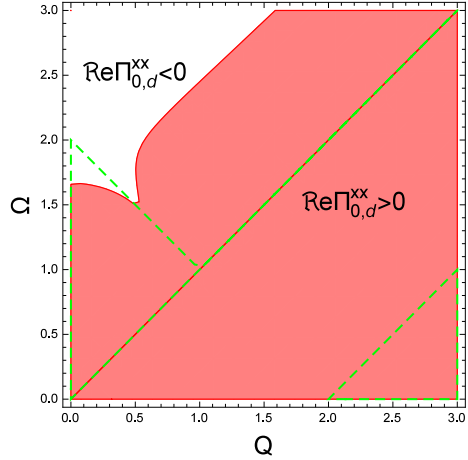}}
         \caption{(Color online) 
	 The possible region for propagation of TE mode in doped graphene. The white (red) region 
	 corresponds to negative (positive) values of$\Re \Pi^{xx}_{0,d}(q\hat{\bs y},\omega)$.
	 Dashed green triangles enclose regions where the imaginary part of this function is zero. The vertical edge of
	 the right triangle further extends if we choose to plot a larger region. On the dashed line along the diagonal
	 also the imaginary part is zero. Horizontal and vertical axis are defined as 
	 dimensionless wave vector $Q=q/k_F$ and energy $\Omega=\hbar \omega/ \ep_F$, respectively.}
         \label{te.fig}
         \end{figure}  
 
  \subsection{TE mode}
  According to equation~\eqref{te-simp}, the TE mode could propagate when the real part of 
$\Pi^{xx}(q\hat {\bs y},\omega)$ is a negative quantity. In the case of undoped graphene as can be seen
from Eq.~\eqref{realundoped}, the current response is determined by $F_0$ which turns out to be a positive 
quantity. Therefore the TE mode does not exist in undoped graphene. 
Therefore let us focus on doped graphene. 

In the case of doped graphene the doping is characterized with $k_F$ corresponding to which
an energy scale $\eps_F=\hbar v_F k_F$ exists. Since the response can be organized in terms of
dimensionless wave vector $Q=q/k_F$ and energy $\Omega=\hbar\omega/\eps_F$, we can construct a
universal diagram for the propagation of TE mode in doped graphene. In Fig.~\ref{te.fig}
the white (red) region indicates where $\Pi^{xx}_{0,d}(q\hat{\bs y},\omega)$ is negative (positive). 
So any possible solution should occur in the white region. In order to avoid damping one must
also search in a region where the imaginary part of $\Pi^{xx}_{0,d}(q\hat{\bs y},\omega)$ is zero.
Regions where the imaginary part is zero are confined to dashed triangles in this figure. Therefore
the undamped TE mode in doped graphene can only exist in the white region inside the
left dotted triangle which shares its side with $\Omega$ axis. 
Indeed at $Q=0$ the possible range of energies corresponding to $1.667<\Omega<2$ which has been
found in Ref.~\cite{mikhailove} is clearly seen in this figure. 
This figure further shows that the region of $(Q,\Omega)$ space where a possible TE mode can
propagate extends up to near $Q\approx 0.5$ and at the same time shrinks by increasing $Q$.
Therefore the TE mode found by Mikhailov~\cite{mikhailove} ceases to exist when the doping
is so low that the $k_F$ becomes close to $\approx2$ times the wave-vector of EM radiation.

\begin{figure}[t]
   \centerline{\includegraphics[width = .5\textwidth] {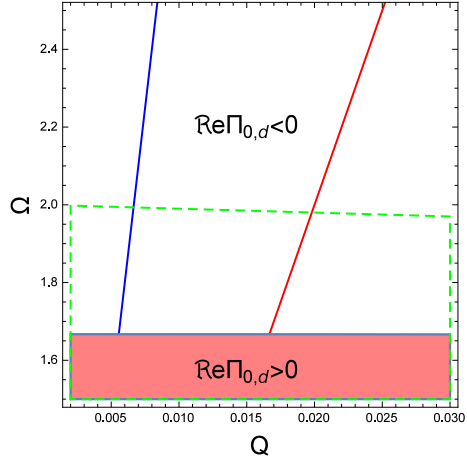}}
    \caption{(Color online) Dispersion of TE mode in doped graphene. The white (red) region corresponds to
    negative (positive) values of $\Re \Pi^{xx}_{0,d}(q\hat{\bs y},\omega)$.
    The dashed green rectangular region is a subset of the triangular region in Fig.~\ref{te.fig} and therefore 
    enclose regions where the imaginary part of this function is zero. The solid blue (red) line corresponds 
    to dispersion of TE mode with $v_F/c=1/300(1/100)$. Horizontal and vertical axis are defined as 
    $Q=q/k_F$ and $\Omega=\hbar \omega/ \ep_F$, respectively.}
    \label{tedisp.fig}
\end{figure}  
 
 \begin{figure}[t]
     \centerline{\includegraphics[width = .5\textwidth] {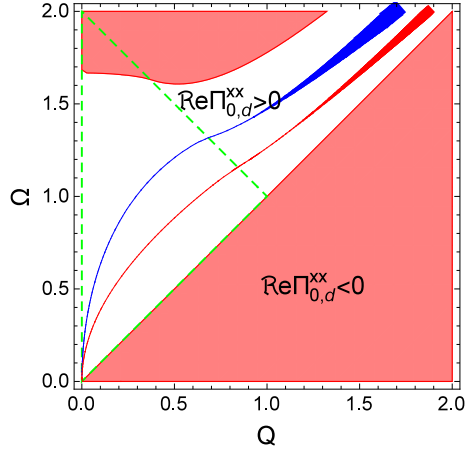}}
    \caption{(Color online) Dispersion of TM mode in doped graphene before dressing the polarization function
    with ladder diagrams. The red (white) area corresponds to the negative (positive) $\Re \Pi^{xx}_{0,d}(q\hat{\bs x},\omega)$. 
Vertical and horizontal axis are $q/k_F$ and $\hbar \omega/ \ep_F$, respectively.
The area enclosed by dashed green line correspond to the region where $\Im \Pi^{xx}_{0,d}(q\hat{\bs x},\omega)$ is zero. 
 The blue (red) line is the dispersion of TM mode with $v_F/c=1/300$ ($1/100$). Outside the green triangular region, TM mode will be damped by the non zero $\Im \Pi^{xx}_{0,d}(q\hat{\bs x},\omega)$.
This damped behavior is represented by broadening which arises from the imaginary part of $Q$. 
{\em Note that for visibility purpose we have magnified the broadening by a factor of $10$. }
}
    \label{tm.fig}
\end{figure}
 So far we have not really solved Eq.~\eqref{te-simp} and have discussed on general grounds the circumstances
under which a {\em possible} solution to this equation may exist. To solve the equation and obtain the dispersion relation, 
one needs the numerical value of the $v_F/c$.
The blue (red) solid dispersion relation in Fig.~\ref{tedisp.fig} denote the TE mode for two different values 
of the parameter $v_F/c=1/300$ ($1/100$). 
The solution to Eq.~\eqref{te-simp} in the white region inside the left
triangle exists only up to a certain value of $Q_{max}$, the value of which is controlled by the numerical value of the
parameter $v_F/c$. For example when $v_F/c=1/300$, we find $Q_{max}\approx 0.0066$ while if $v_F/c=1/100$ (due to enhancement
of the Fermi velocity in ultralow doped graphene) we have $Q_{max}\approx0.02$. Beyond this points there are no real $(Q,\Omega)$ solutions to the TE mode in graphene. 
           
  \subsection{\label{sec:level1}TM mode}
The propagation of TM mode is governed by Eq.~\eqref{tm-simp} which includes the longitudinal current correlation 
$\Pi^{xx}(q\hat{\boldsymbol{x}},\omega)$. The solutions to TM mode dispersion can exist only when real part of this quantity
is positive. In the case of undoped graphene again we need Eq.~\eqref{realundoped} from which the longitudinal (i.e., $\theta=0$) 
part of the response turns out to be negative and hence forbids the propagation of TM mode in undoped graphene.

In Fig.~\ref{tm.fig} we have plotted the sign structure of the $\Pi^{xx}_{0,d}(q\hat{\boldsymbol{x}},\omega)$. For
the positive (negative) real part where the TM mode can (can not) have a solution we have used the white (red) color.
Again the region where the imaginary part of this function is zero is enclosed by dashed green line. Therefore the
intersection of white region and dashed triangle is a region where undamped TM mode can exist, while the white region 
outside the dashed green triangle is the region where the TM mode is damped by dissipating energy to particle-hole
pairs. In the optical limits i.e., $q=0$ this white window is $\omega\lesssim 1.667$ and by increasing 
momentum $q$ it will be slightly more restricted as in Fig.~\ref{tm.fig}.
The blue and red solid lines are the solutions of Eq.~\eqref{tm-simp} for representative value of 
$v_F/c=1/300$  and  $v_F/c=1/100$. 
The group velocity of TM mode is larger when $v_F/c$ is larger at very small momenta. 
When these dispersion relation cross the boundary of the dashed triangle the imaginary part of longitudinal
current response will be nonzero which causes the TM mode to dissipate energy to particle-hole pairs and acquire 
damping. 
The broadening of the dispersion indicated in Fig.~\ref{tm.fig} is a measure of the
imaginary part $Q^\dprime$ of the wave vector $Q$. In appendix D we show how the imaginary
part $Q^\dprime$ is related to the imaginary part of $\Pi_{0,d}^{xx}$. See Eq.~\eqref{tmbroad.eqn}. 
Note that in this figure to clearly represent the damping we have multiplied the broadening by a factor of $10$.

  \section{\label{sec:level1}Dressing the polarization by ladder diagrams}
So far we have shown that equations~\eqref{te-simp} and~\eqref{tm-simp} give the 
dispersion of TE and TM modes in terms of transverse and longitudinal parts of the
two-current tensor. In this section we would like to study which one of the above responses
receives corrections from the ladder diagrams generated by the Hubbard interaction. For this purpose we start by the
continuity equation which is given by,
\be
  \label{contin}
  \div \boldsymbol{\hat{j}}(\boldsymbol{r},t)=\partial_t \hat{\rho} (\boldsymbol{r},t),
\ee
and relates current and density operator. As a consequence of this, current and density correlation function~\cite{principi,scholz} 
are related by,
\be
  \label{curren-density}
  \chi(q,\omega)=\frac{q}{\omega^2} [\hat{\bs j}_x,\hat{\rho}_{-q}] (q\bs{\hat{x}},\omega)+ 
  \frac{q^2}{\omega^2} \Pi^{x,x} (q\bs{\hat{x}},\omega).
\ee
In the right hand side of this equation first term turns out to be a real quantity which is proportional to cutoff parameter and will be 
canceled by similar cutoff term in the second term. Therefore the longitudinal current response function has a direct relation with 
density correlation function, 
\be
   \chi(q,\omega)= \frac{q^2}{\omega^2} \Pi^{xx} (q\bs{\hat{x}},\omega).
   \label{fdc}
\ee
Therefore, if any correction such as inclusion of ladder diagrams causes some changes in the 
behavior of (electric charge) density response, it will directly 
affect the longitudinal current response via Eq.~\eqref{fdc}. Normally used approximation
which corresponds to random phase approximation is to use $\chi=\chi_0$ which corresponds to the
first term in Fig. 1b. However, as will be shown shortly, the inclusion of ladder diagrams
will totally change the structure of $\chi$, and hence will drastically affect the propagation of TM 
mode in graphene. 
This will in turn modify {\em only} the propagation of TM mode. 
Therefore in principle the TM mode receives corrections from the inclusion of dressing of the ladders
by Stoner PH processes, while the TE mode is not affected by such diagrams as they ultimately arise from
a density-density (Hubbard) interaction.
Only Thirring type (current-current) interactions are able to modify the TE mode which maybe relevant for
Thirring matter, but not to graphene as an example of Dirac systems on which we are focused in this paper. 

Let us point out that the particle-hole pairs propagating along the rungs of ladder have two
options when they enter a dashed interaction line that represents the Hubbard interaction:
They can either flip their spins for which the interaction will chagne sign to $-U$; or they 
can keep going with no spin flip~\cite{scalapino,senechal,takimoto,wen,mishchenko} for whith the Hubbard term 
keeps its original sign $+U$. 
If one tries to construct a (particle-number) conserving approximation
in the sense of Baym and Kadanoff~\cite{BaymKadanoff}, both sets of diagrams are relevant and none of them can be neglected
in the interest of the other. This is the essence of fluctuation-exchange approximation (FLEX) which in a sense Fierz decomposes
the (short range) Coulomb interaction into various channels such as singlet or triplet PH channels, and similar particle-particle channels.
For any model Hamiltonian, the properties of the system decide fluctuations in which channel are to be enhanced or suppressed.
The nature of particle-hole continuum in graphene is such that provides a unique opportunity of which the spin-flip fluctuations 
can take advantage and develop a coherent pole which indicates that they bind into triplet bound states of particle-hole pairs 
below the continuum of free particle-hole excitations~\cite{Jafari2002,Jafari2004,Jafari2012,Jafari}.  
The very same ladder summation mechanism when accounted for in the EM response of the system, generates a singularity
in the polarization function that arises from proliferation of Stoner PH excitations along the ladder rung. 
Therefore the major role is played by the spin-flip processes that run across the rungs of the ladder. 

Putting the above words on a formal language has been already done in the literature and is
known as the fluctuation exchange (FLEX) approximation which is a 
natural language to address the role of separate channels for fluctuations of various quantities~\cite{Scalalpinoann,scalapino}.

For short range interactions where Hubbard $U$ is independent of momentum, the ladder series in Fig. 1b can be easily 
summed. This has been already done in the FLEX approximation in a more complete form that takes both spin-flip and
non-spin-flip processes along the ladder rung. 
Since PH fluctuations in both spin-flip and non-spin-flip channel will ultimately lead to fluctuations of the electric charge,
the effective interaction determining the dielectric properties will have contributions from both channels. The
part of effective interaction that is attached to two bare interaction lines and defines the effective 
charge polarization, in the FLEX approximation is given by~\cite{scalapino,senechal,takimoto,takimoto1,yamada}, 
\begin{align}
   \chi^{\rm eff}_{\rm FLEX}(q,\omega) = &
   \frac{3}{2}  \bigg[\frac{\chi_0(q,\omega)}{1+U\chi_0(q,\omega)}-\chi_0(q,\omega) \bigg]+\nn\\&
   \frac{1}{2}    \bigg[\frac{\chi_0(q,\omega)}{1-U\chi_0(q,\omega)}-\chi_0(q,\omega)\bigg]+\nn\\&
   \chi_0(q,\omega)  
 \label{flex}
\end{align}
where $\chi_0(\bsq,\omega)$ is density response function of non-interacting system without the spin degeneracy.
Note that despite its RPA-{\em looking} form, the above formula is related to ladder summation. It is only
due to momentum independence of Hubbard $U$ interaction that the ladder summation acquires such a simple form. 
The first and second lines represent the contribution of spin-flip and non-spin-flip channels, respectively.
Note that as expected from our simple argument in the introduction, the sign of interaction in the spin-flip
channel differes from the non-spin-flip channel. As a quick check, the terms in the brackets vanish when we set $U=0$
and one is left with empty bubble $\chi_0$ only. 
As we will see shortly, the sign change $U\to -U$ in the denominator of the spin-flip channel,
will make the $\chi^{\rm eff}_{\rm FLEX}$ positive and therefore Eq.~\eqref{tm-simp} can have 
a solution in undoped graphene.


As pointed out earlier, the longitudinal (density) interactions of the Coulomb type are not
going to affect the propagation of TE mode. So we focus on the TM mode and consider the 
role of triplet fluctuations on propagation of the TM mode in graphene. Let us start by 
the undoped graphene first.

\subsection{\label{sec:level2}Undoped graphene}
As we pointed out earlier, propagation of TM mode in undoped graphene is not possible as a result of negative current response function.
When the contributions from spin flip channel of the ladder diagrams that corresponds to the first line of Eq.~\eqref{flex} 
is taken into account, the negative sign of $\chi_0$ below the PH continuum can generate a pole in the ladder summation of 
spin-flip processes. This gives rise to a positive $\chi^{\rm eff}_{\rm FLEX}$ 
which in turn can give rise to propagating solutions in Eq.~\eqref{tm-simp}.
The ladder summation with long-ranged (momentum dependent) interactions can be performed under very stringent conditions
which is valid for $\omega\approx qv_F $~\cite{mishchenko}. In this case, a singularity in the density response can
be produced by the ladder diagrams and gives rise to a {\em damped} solution in Eq.\eqref{tm-simp}. 
We find that this solution is $\omega\approx 1.15 qv_F$ which lies above the $\omega=qv_F $ line where the density of 
free PH pairs is non-zero and therefore becomes damped. Inclusion of ladder diagrams reduces the imaginary part of 
$\chi$ and thereby reduces the damping, but this is not sufficient. A better situation can be generated with
short range interactions (which is already strong enough in graphene) that enhances spin fluctuations 
This enhancement is precisely encoded in the denominator of the first line of Eq.~\eqref{flex} and is known as Stoner enhancement~\cite{Moriya}.

\begin{figure}[t]
    \centerline{\includegraphics[width = .5\textwidth] {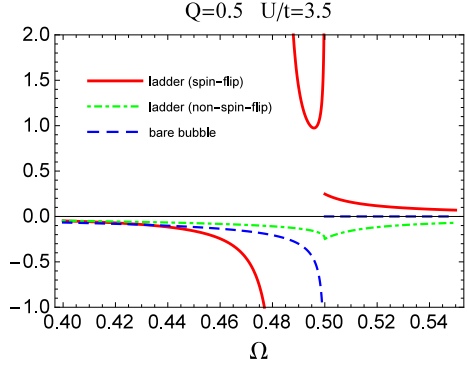}}
    \caption{ (Color line) Separate contribution of bare bubble (blue dashed line) corresponidng to no ladder correction, 
    spin-flip ladder diagrams (red solid line) and non-spin-flip ladders (green dotted dashed line). These terms when 
    added up give the effective charge polarization in Eq.~\eqref{flex} for $U/t=3.5$ and $Q=0.5$ in undoped graphene.
    Note that in undoped graphene we use the following definition $Q=\hbar q v_F/ t$ horizontal axis is $\Omega=\hbar \omega/ t$. 
    As can be seen only the red line allows $\chi^{\rm eff}$ to become positive and hence the TM mode equation can develope possible solutions.
}
    \label{undopedflex.fig}
\end{figure}

To see the structure of sings, in Fig.~\ref{undopedflex.fig}, we have plotted the charge polarization arising from
various terms of Eq.~\eqref{flex}. The bare polarization (blue dashed line) corresponds to the first term in the
right hand side of Fig.~\ref{bubbles.fig}b and is given in third line of Eq.~\eqref{flex}. 
The first line of this equation corresponds to the rest of diagrams in Fig.~\ref{bubbles.fig}b and is
denoted by red solid line in Fig.~\ref{undopedflex.fig}. Finally the second line of Eq.~\eqref{flex}
corresponds to ladder diagrams along the rung of which spins do not flip are denoted by green dashed line.
This figure is produced for $U/t=3.5$ and $Q=0.5$ and demonstrates how the inclusion of the ladder diagrams and the spin-flip fluctuations
encoded in these diagrams can make the resulting $\chi^{\rm eff}$ positive and hence lead to a solution for the TM mode Eq.~\eqref{tm-simp}. 
This figure clearly shows that the dominant effect of ladder diagram resummation comes from spin-flip excitations 
and therefore the resulting divergence is the Stoner enhancement.

Now that we are convinced about the special role of Stoner enhancement in making the $\chi^{\rm eff}$ positive, and
hence providing a chance for a solution to the TM mode Eq.~\eqref{tm-simp}, let us study the dependence of
solutions to the Hubbard parameter $u=U/t$ which controls the strength of spin-fluctuations. To this end,  
in Eq.~\eqref{tm-simp} we place the $\Pi^{xx}$ in one side and the rest in the other side of equation
and in $\Pi^{xx}$ we take full account of all channels in Eq.~\eqref{flex}. 

\begin{figure}[t]
  \begin{tabular}{cc}
   \subfloat (a){\includegraphics [width = .40\textwidth]{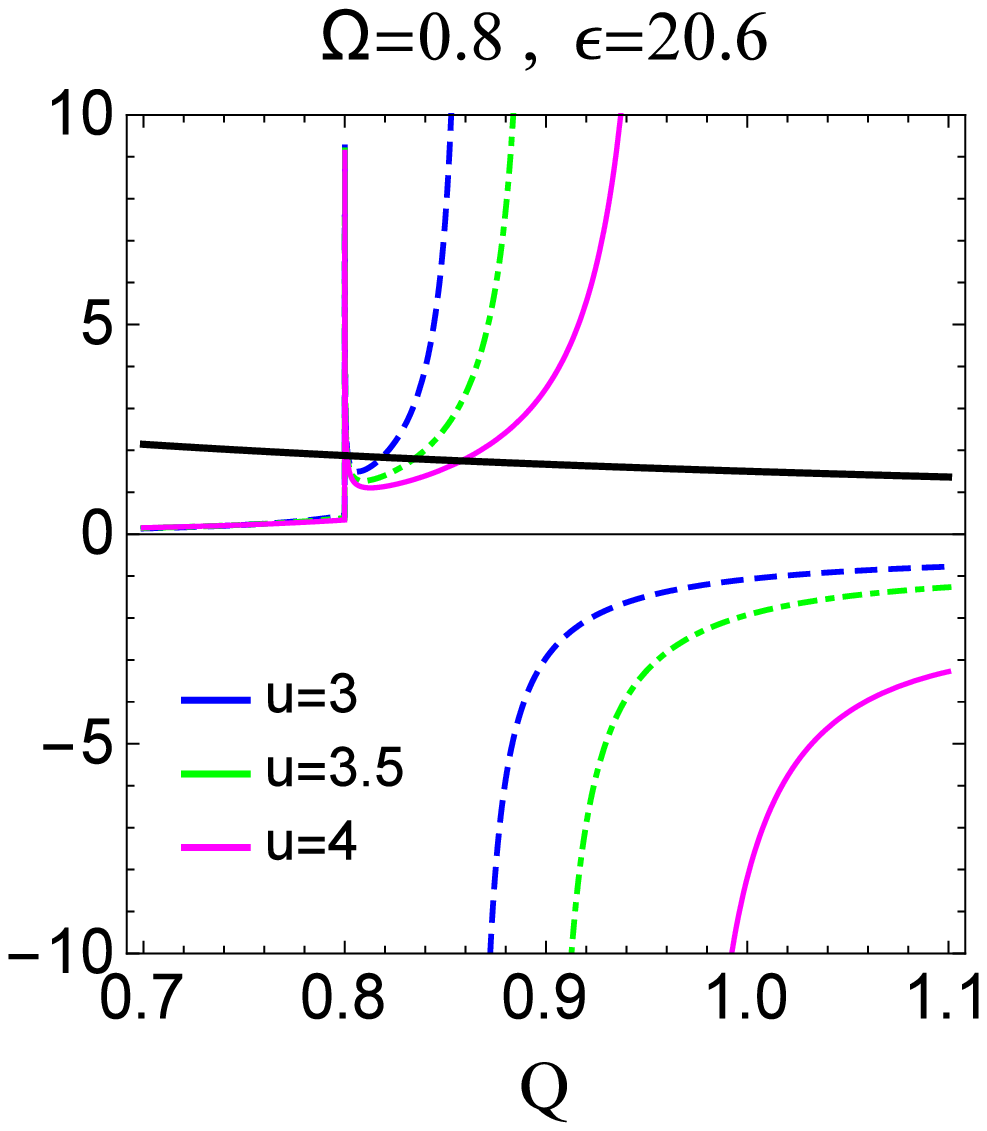}} &
   \subfloat (b){\includegraphics[width = .40\textwidth] {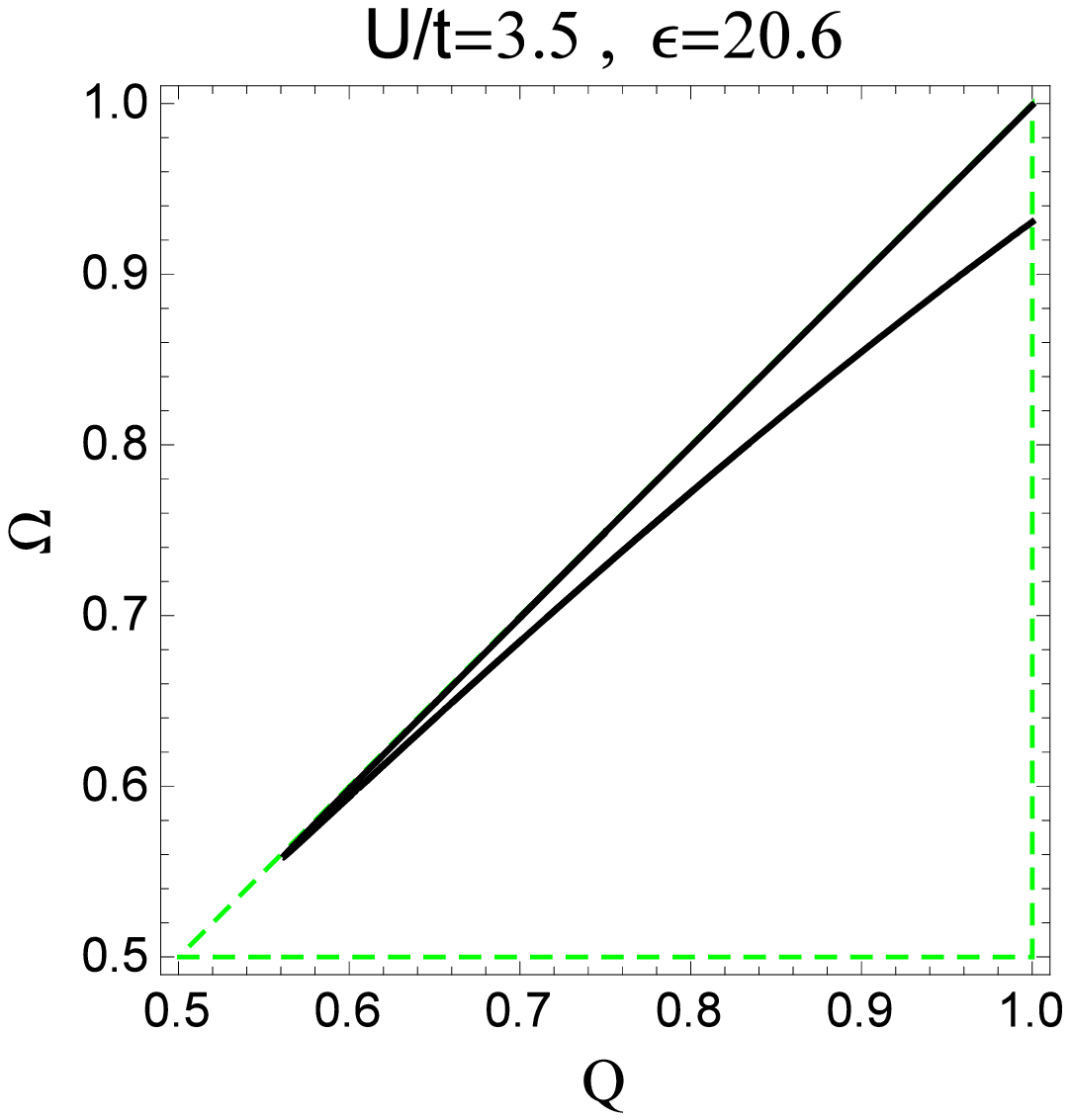}}
   \end{tabular}
    \caption{ (Color line) (a) Solutions for TM modes in undoped graphene within the ladder resummation approximation for 
    different value of repulsive Hubbard parameter $u=U/t$. Here grpahene layer is surrounded by dielectric 
    constant $\ep\sim20.6$. In panel (a) the thin lines (blue dashed, green dotted dashed, pink solid) are plot of 
    $\Pi^{xx}$ within the ladder approximation, Eq.~\eqref{flex} and the black (thick solid) line is plot of other side 
    of equation obtained from re-arranging Eq.~\eqref{tm-simp}. 
    Panel (b) depicts the dispersion of TM modes for indicated values of $U$ and $\epsilon$. The axes are defined
    by $Q=\hbar q v_F/t$ and $\Omega=\hbar \omega/t$.
}
    \label{undopedflex1.fig}
\end{figure}

In the left panel of Fig.~\ref{undopedflex1.fig} we have plotted the the $\Pi^{xx}$ for various values of $u=U/t$ and assumed
that the surrounding medium is such that the effective dielectric constant is $\epsilon\sim 20$.
In the right panel of this figure we have extracted the dispersion of TM mode for $U/t=3.5$ and $\epsilon=20.6$.
As can be seen from panel (a) there are always two solutions, one is immediately below the PH continuum, 
and the other is well below the PH continuum and therefore protected from damping. Note how 
larger Hubbard parameter increases the separation of TM mode solution from the PH continuum and hence
leads to better protection from damping. In panel (b) of this figure, for a fixed $u=U/t=3.5$ adopted
from Ref.~\cite{wehling} for graphene, and for surrounding medium dielectric constant of $\epsilon=20.6$
we plot the dispersion of TM modes in black. 
Both TM dispersions start at a threshold wave vector.
Increasing the effective dielectric constant of the surrounding medium brings this threshold 
wave vector to smaller values. 

 \subsection{\label{sec:level3}Extrinsic graphene}
So far we have seen that the modifications of the charge polarization by spin-flip PH fluctuations 
can in principle give rise to a TM mode solution in undoped graphene which can never happen if 
no spin-flip fluctuations are taken into account. Now let us turn our attention to dopped graphene.
In doped graphene, the TM modes are sustained in the form of plasmon excitations of the chiral
electron gas~\cite{hwang}. Now the question is whether the inclusion of the spin-flip ladder diagrams will
affect TM mode or not? We will shortly see that the generic role of itinerant spin fluctuations in graphene is to
lower the energies of TM modes.

\begin{figure}[t]
     \begin{tabular}{cc}
       \subfloat (a){\includegraphics [width = .40\textwidth]{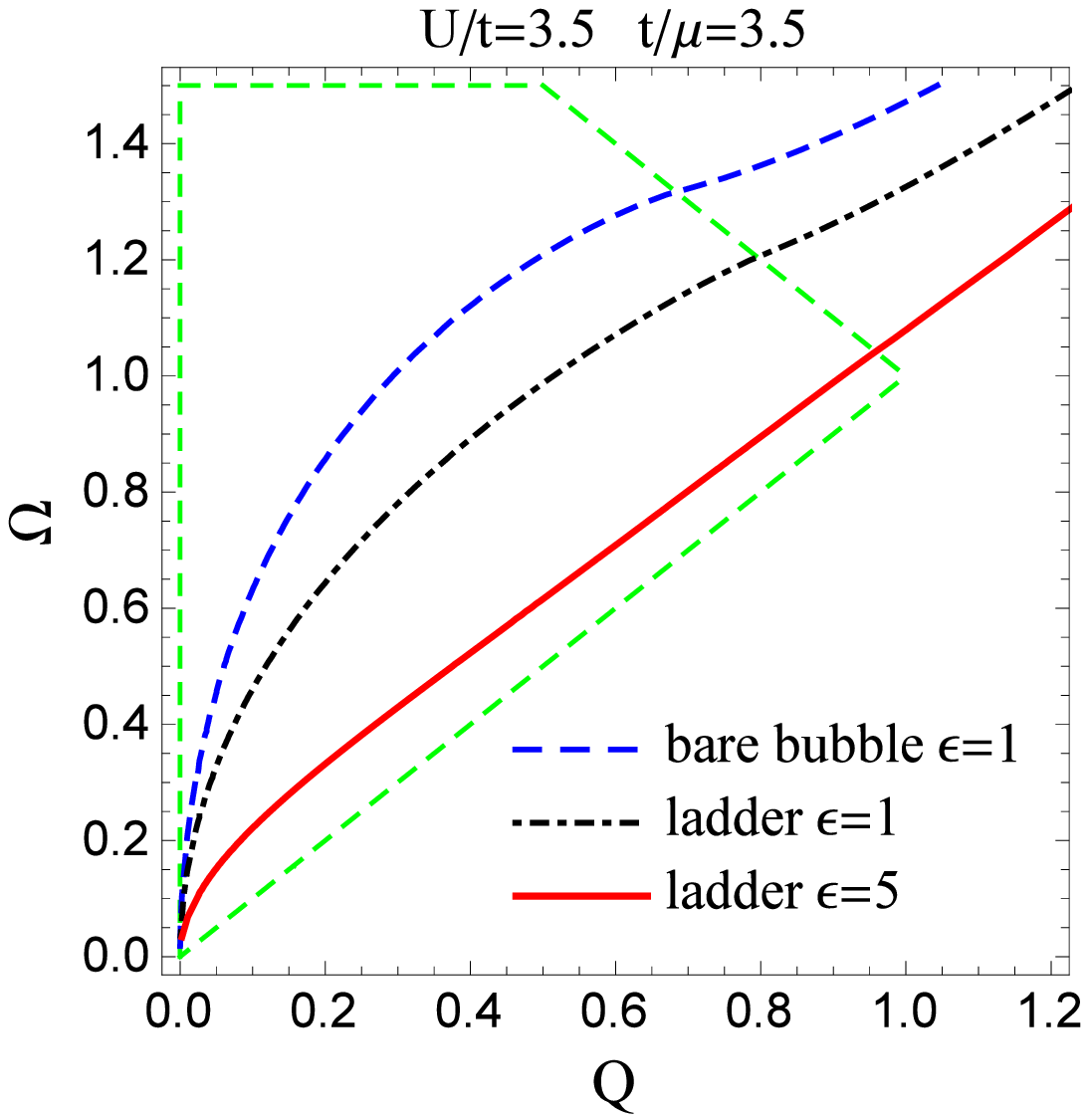}} &
       \subfloat (b){\includegraphics[width = .40\textwidth] {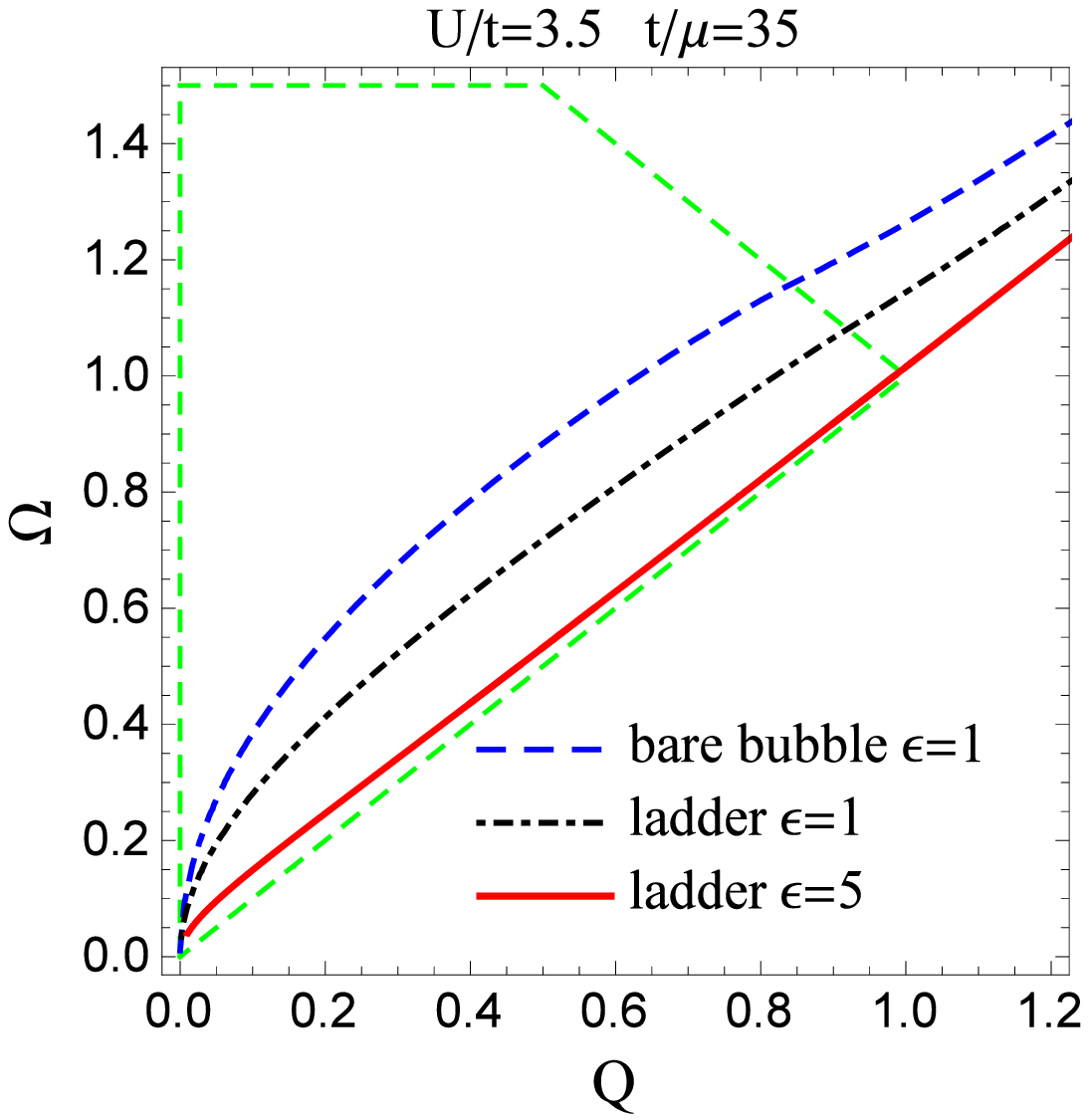}}
       \end{tabular}
       \caption{ (Color line) Propagation of TM mode in doped grpahene. We compare the bare (blue dashed), with full FLEX 
charge polarizations as indicated in the legend for medium dielectric constants of $1$ (black, dotted dashed line) and $5$ (red, solid line). 
Panel (a) and (b) show dispersion of TM mode in typically doped ($v_F/c=1/300$ , $t/\mu=3.5$) and 
ultra-low dopped ($v_F/c=1/100$ , $t/\mu=35$) graphene. Here horizontal and vertical axes 
are  $Q=\hbar q v_F/ t$ and  $\Omega=\hbar \omega/ t$ respectively where $\mu$ is Fermi energy, $t$ hopping amplitude.}
       \label{dopedflex1.fig}
   \end{figure}

To elucidate the peculiar role of spin fluctuations on the TM mode of doped graphene, in Fig.~\ref{dopedflex1.fig}
we plot the dispersion of the TM mode in doped graphene for three different
situations: bare (blue dashed line) corresponding to no ladder corrections,  and full ladder approximation
of Eq.~\eqref{flex}. 
The black (dotted dashed) line corresponds to the
effective dielectric constant $\epsilon=1$ and the red (solid) line corresponds to $\epsilon=5$ as indicated in the
legend in both panels. The value of Hubbard $U$ in both panels is $U/t=3.5$.
In panel (a) is for typically doped graphene with $t/\mu=3.5$ , $v_F/c=1/300$ while the
panel (b) is for ultra-low doped graphene $t/\mu=35$, $v_F/c=1/100$.
As can be seen in all cases the role of triplet PH fluctuations encoded in the ladder summation is to 
lower the energy of TM mode. If graphene is surrounded with an dielectric constant $5$, the dispersion of the TM mode will 
come closer to the $\Omega=Q$ line.
The ultra-low doped graphene shows the same features but with differences which arises of smaller Fermi surface and bigger Fermi velocity. 
The dispersive mode inside (outside) of the area with green dashed line is undamped (damped).
 The common feature of all dispersions is that once the ladder resummation is included, the energy of the
TM mode will be lowered. 

An interesting observation in Fig.~\ref{dopedflex1.fig} is that when the Fermi velocity is 
parametrically larger (which can be achieved in ultra-low doping by renormalization~\cite{ultralow}), 
inclusion of the spin-flip particle-hole fluctuations gives rise to a branch of TM mode which almost
disperses at the Fermi velocity. The conclusion is that 
the total spin of the particle-hole pairs has a significant effect on the propagation of the TM mode,
and {\em the TM mode can take advantage of the minus sign in the triplet channel of Eq.~\eqref{flex}
to lower its energy}. This can be interpreted as follows: An incoming photon propagator
creates an electron-hole pair. This Electron-hole pair due to strong spin fluctuations
which can be thought of as an effective boson that mediates spin-flip across the ladder,
gives rise to a form of dressing which makes $\chi{\rm eff}$ positive and hence can 
furnish a solution to Eq.~\eqref{tm-simp}. At the end of this process, the end particle-hole
pairs recombine to emit a photon.

\section{\label{sec:level1}Summary and conclusion}
In this work we have investigated the role of Stoner enhancement which is generated by 
strong enough short-range interactions -- and is formalized as series of ladder diagrams
as in Fig.~\ref{bubbles.fig} -- on the propagation of electromagnetic modes. 
An incident light generates a PH pair. This pair is then
resonantly scattered as shown in the ladder summation in Fig.~\ref{bubbles.fig}.

We have chosen to demonstrate this effect in graphene, because: (i) Graphene has large
enough Hubbard $U\sim 10$ eV. (ii) When doped it supports an interesting TE mode. 
(iii) The nature of PH continuum is such that the spin-flip fluctuations can be resummed
into a coherent pole that lies {\em outside} the PH continuum. Under such a suitable circumstances, the
inclusion of ladder diagrams has its most drastic effect in the spin-flip channel as
in this channel the sign of $U$ is reversed: $U\to -U$. This sign reversal gives rise
to a resonance in undoped graphene where it gives rise to a branch of TM mode very well 
separated from the PH continuum. This minus sign picked from fermionic anti-commutations on going 
from non-spin-flip to spin-flip channel provides a unique chance to get a solution to 
Eq.~\eqref{tm-simp} which is otherwise (i.e. without ladder corrections) impossible. 
{\em Realization of this effect requires high dielectric surrounding medium.}

In doped graphene an intrinsic momentum scale $k_F$ and energy scale $\epsilon_F$
emerge and hence physical properties are functions of the ratio $\bsq/k_F$ and $\hbar\omega/\epsilon_F$.
Possibility of ultralow doping in graphene, provides access to finite $\bsq$ behavior of the optical response. 
Inclusion of ladder corrections slightly modifies the TM modes of doped graphene by
generically lowering their energies. In the ultralow doped case where the Fermi velocity
is parametrically large, this effects becomes substantial. 
This lowering of energy of the TM mode can be traced back to the minus sign picked up
in the spin-flip channel. 
The ladder corrections do not affect the TE modes~\cite{mikhailove} of the doped graphene. 

What do the spin-flip particle-hole fluctuations do in conducting states other than graphene, e.g.
in a normal metal? The essential property of graphene is reflected in the 
existence of a region {\em below} the continuum of PH excitations.
The particle-hole continuum of normal conductors (i.e. Fermi liquids) does not allow for such windows below the continuum.
Such a window below the continuum when combined with relatively large on-site Coulomb repulsion $U/t\sim 3.5$~\cite{wehling}
can enhance the spin fluctuations. But since the spins are not localized, spin-1 fluctuations also 
contribute to the charge polarization, albeit with effectively reversed sign of Hubbard $U$.
In normal conductors when the triplet fluctuations want to bring down the energy of the
TM mode, the mode sinks more into the continuum of free particle-holes and gets quickly damping. 
Moreover, in most normal conductors the screening is very effective, and conduction bands are
very wide, such that the ratio of the Hubbard $U$ and hopping is not large. 
Therefore the chiral nature of fermions in graphene along with the triplet fluctuations arising from Hubbard
interactions join hands to give a unique chance for the propagation of TM mode in undoped graphene.

The nature of PH continuum in higly oriented pyrolytic graphite 
-- above the energy scale of $\sim 50$ meV related to inter-layer hopping between the constituting 
graphene layers -- is quite similar to graphene~\cite{Jafari2002}. Therefore the same effects are expected for
graphite as well. Propagation of circularly polarized EM radtion with graphene/graphite 
may also receive corrections from spin fluctuations. This might have relevance to
natural birefringence observed in a closely related material, graphite~\cite{Mertins,Grimmer}.

\appendix
 \section{\label{sec:level1} Current response function}
 Electromagnetic response of graphene is characterized by the tensor form of current correlation function Eq.~\eqref{pikubo.eqn} in the terms of form factor elements. If we substitute $\phi_k$ by $\phi_k+\theta$ in current correlation function Eq.~\eqref{pikubo.eqn}, we obtain,
 \begin{align}
   \label{cos}
   \cos(\phi_k +\phi_{k+q})=\frac{ k\cos2(\phi_k+\theta)+q\cos(\phi_k+2\theta)}{|\bs k+\bs q|},\\
  \label{sin}
   \sin(\phi_k+\phi_{k+q})=\frac{ k\sin2(\phi_k+\theta)+q\sin(\phi_k+2\theta)}{|\bs k+\bs q|},
 \end{align}
  where
\be
   |\bs k+\bs q|= \sqrt{k^2+q^2+2kq\cos\phi}~.
\ee
The advantage of point of this change of variable is that allows
us to represent each element of tensor current correlation function in terms of one function, e.g., $ \Pi^{x,x}(\boldsymbol{q},\omega)$. By using properties of  this transformation the new representation of  tensor current correlation function in terms of $ \Pi^{x,x}(\boldsymbol{q},\omega)$ will be given by  Eq.~\eqref{matrixform}. In what follow, we use this representation of tensor current correlation function.

We better define the difference between the corresponding
response between doped and undoped case for $xx$ component of current response tensor as,
\bearr
\label{app-deltapi}
 \Delta \Pi^{x,x}(\boldsymbol{q},i\omega_n)&&= \Pi^{x,x}_d(\boldsymbol{q},i\omega_n) - \Pi^{x,x}_u(\boldsymbol{q},i\omega_n)\nn\\&& =  \frac{g_s g_v v_F^2}{4\pi^2 \hbar}   \int_{0}^{k_f} k dk d\phi \nn\\&&\times [P(\boldsymbol{k},\boldsymbol{q},i\omega_n) + P^*(\boldsymbol{k},\boldsymbol{q},i\omega_n)],
 \eearr
where the subscripts $u,d$ stand for undoped and doped, respectively, and 
 \bearr
 P(\bsk,\bsq,i\omega_n) = && \frac{(i\omega_n + v_F k)}{(i\omega_n+ v_F k)^2 -v_F^2 |\bsk+\bsq|^2} \nn\\&&+ \frac{v_F k \cos2(\phi+\theta) }{(i\omega_n+ v_F k)^2 -v_F^2 |\bsk+\bsq|^2} \nn\\&&+\frac{v_F q \cos(\phi+2\theta)}{(i\omega_n+ v_F k)^2 -v_F^2 |\bsk+\bsq|^2},
 \eearr
which has a nice quadratic expressions in the denominator. Here, $i\omega_n$ is $\omega+i\eta$ which points to  the complex  frequency domain with infinitesimal imaginary part $\eta$. One can easily see the terms which include $\sin 2\phi$ or $\sin \phi$ don't have any contribution hence, in the following calculation we eliminate them in the representation of $ P(\bsk,\bsq,i\omega_n)$ and it's conjugate and proceed our calculation by using complex integration.

 \subsection{\label{sec:level1} real part}
  As a result of substituting $i\omega_n$ by $\omega$, 
 the real part of response function is defined by
 \bearr
  \Delta \Pi^{xx}(\boldsymbol{q},\omega)&&=   \frac{g_s g_v v_F^2}{4\pi^2\hbar }  \int_{0}^{k_F} k dk d\phi \nn\\&&\times [P(\bsk,\bsq,\omega) + P(\bsk,\bsq,-\omega)]. \label{realpart}
  \eearr
  Here, $P(\bsk,\bsq,-\omega)$ is a real function. For this purpose we define $z=e ^{i\phi}$ and $\Delta=(v_f^2 q^2 -\omega^2-2 \omega v_f k)/(2v_f^2 kq)$ in order to do angular integration,
  \bearr
 \int d\phi   P(\boldsymbol{k},\boldsymbol{q},\omega)=\frac{-i}{2v_F^2 kq} \oint \frac{dz}{z^2} \frac{B_z(\bs{k},q,\omega)}{z^2+2\Delta z+1} 
  \eearr
where,
   \bearr
  B_z(\bs{k},q,\omega)=&&(\omega +2v_F k \sin^2 \theta) z^2+2v_Fk  \cos2\theta(z^2+1)^2+\nn\\&& v_Fq \cos2\theta z (z^2+1).
   \eearr
The roots of denominator are: a second order root at $z_0=0$ and two first order roots at $z_{\pm}=-\Delta\pm\sqrt{\Delta^2-1}$ when $ \Delta\geq1 $ and  $z'_{\pm}=-\Delta \pm i\sqrt{1-\Delta^2}$ when $ \Delta<1 $. Among these roots, whichever placed within the unit circle contribute to the integration. We can summarize result of integration by,
 \be
 \label{step1}
 \int d\phi   P(\boldsymbol{k},\boldsymbol{q},\omega)=Res(z_0)+ \text{sgn}(\Delta) Res(z_+)  \Theta(|\Delta|-1)
\ee
 with 
 \bearr
 Res(z_0)&&=\frac{2\pi \cos2\theta}{v_F q}\omega(\omega+2v_Fk),
 \eearr
and 
 \bearr
 Res(z_+)=&&\frac{2\pi}{\sqrt{\Delta^2-1}}
 \bigg[ 2(\omega+v_Fk) \cos^2 \theta- \nn\\&& \cos2\theta \frac{(\omega+2v_Fk)^2((v_Fq)^2-\omega^2)}{2v_F^3kq^2} \bigg].
 \eearr
 Here, we refuse to include  residue in  $z'_{\pm}$ in Eq.~\ref{step1}, because when $|\Delta|<1$ both of roots simultaneity  lie in the unit circle and due to $Res(z'_-)=-Res(z'_+)$ their contribution in Eq.~\ref{step1} will be cancel each other. 
 Meanwhile the residue in $z_-$ is equal to $-Res(z_+)$. The symmetry of real response function to  frequency can easily seen in Eq.~\ref{realpart}. Therefor it is just sufficient to proceed our calculation for positive $\omega$ and 
 then pick the even part of $\omega$ dependence at the end. 
 The final process  is integration over momentum $k$. The first part of integration ($Res(z_0)$) is proportional to $\omega^2+v_F k \omega$ from which the second term will be dropped upon  $\omega\rightarrow -\omega$ in $P(\boldsymbol{k},\boldsymbol{q},-\omega)$. Finally the contribution of this root is $-\frac{g_s g_v }{2\pi } \ep_F (\omega/\hbar v_Fq)^2\cos2\theta$. 
 The result of integration of $Res(z_+)$ depends on the relation between $\omega$ and $v_F q$ as follows,
 \bearr
  &&  -\frac{g_s g_v  }{2\pi }  \frac{ \ep_F \omega^2}{(\hbar v_Fq)^2} \cos2\theta +\frac{g_s g_v s(\Delta)}{16 \pi \hbar} \Theta(v_Fq-|\omega|)  \nn\\ && \times
   \left[C_1G\left( x\right)+C_2 \sin^{-1}\left( x\right)\right]^X_0 \equiv \left[F_1^{\pm \omega}(x)\right]^X_0
 \eearr
 
 \bearr
 &&
    -\frac{g_s g_v  }{2\pi }  \frac{\ep_F \omega^2}{(\hbar v_Fq)^2} \cos2\theta +\frac{g_s g_v \text{s}(\Delta)}{16 \pi \hbar} \Theta(|\omega|-v_F q) \nn\\ && \times\left[
    -C_1 G\left( x \right) + C_2 \cosh\left( x\right)\right]^X_0 \equiv \left[F_2^{\pm \omega }(x)\right]^X_0 
 \eearr
 where 
 \be
 C_1=\frac{\omega^2\cos^2\theta -\left(\omega^2-\left( v_F q\right)^2\right)\sin^2\theta}{\sqrt{\left|\omega^2-\left( v_F q\right)^2\right|}},
 \label{c1.eqn}
 \ee
 \be
 C_2=\frac{\omega^2\cos^2\theta +\left(\omega^2-\left(v_F q\right)^2\right)\sin^2\theta}{\sqrt{\left|\omega^2-\left( v_F q\right )^2 \right|}},
 \label{c2.eqn}
 \ee
 \be
 G\left( x\right)=\left(\frac{2x v_Fk_F+\omega }{ v_F q} \right) \sqrt{\left|\left(\frac{ 2x v_Fk_F+\omega }{ v_F q}\right) ^2-1\right|},
  \label{g.eqn}
  \ee
 \be
  \cosh \left(x\right)= ln\left[\left| \frac{2 x v_Fk_F+\omega}{ v_F q}\right|+\sqrt{\left|\left(\frac{ 2 x v_Fk_F+\omega}{ v_F q} \right)^2-1\right|}\right],
   \label{cosh.eqn}
 \ee
 \be
  \sin^{-1}\left( x\right)=\sin ^{-1}\left(\frac{ 2 x v_Fk_F+\omega}{ v_F q} \right).
   \label{sin-1.eqn}
 \ee
 Here $\pm \omega$ in both functions refer to the integration over $P(\boldsymbol{k},\boldsymbol{q},\pm \omega)$ and the function $s(\Delta)$ is basically $sgn(\Delta)\Theta\left(|\Delta|-1\right)$, and upper limit $X$ are functions of $q,\omega,k_F$. Before finding the integration range, note that the functional form of the above expression at $x=0$ is purely odd with respect to $\omega$ and therefore drops out by adding the $\omega \to -\omega$. We therefore need to carefully determine the upper limit $X$ which was shown in  Fig.~(\ref{partitioning.fig}). 
  \subsection{\label{sec:level1} Imaginary part}
In Eq.~\ref{app-deltapi}, if we consider imaginary contribution of  $P^{*}(\boldsymbol{k},\boldsymbol{q},i\omega_n)$ and $P(\boldsymbol{k},\boldsymbol{q},i\omega_n)$, imaginary part of $\Delta \Pi$ 
will be derived.
Again we focus on first term of this representation i.e., $P(\boldsymbol{k},\boldsymbol{q},\pm \omega)$ and repeat similar calculation for other term. As a first step we profit by presence of infinitesimal imaginary part in $\eta$ in $\omega+i\eta$ which causes a delta function of the form
   \bearr
   \Im [P(\boldsymbol{k},\boldsymbol{q},\omega)]\equiv\frac{M(\bs{k},\bsq,\omega)}{4(\omega+v_F k)^2} \delta(\xi),
   \eearr 
where,
    \bearr
   M(\bs{k},\bsq,\omega)=&&\omega^2+2 \omega v_F k -v_F^2 q^2-2 v_F^2 kq \cos \phi \nn\\&&-2(\omega +2v_F k \sin^2 \theta+2v_Fk  \cos2\theta \cos ^2\phi\nn\\&&+ v_Fq \cos2\theta  \cos \phi)(\omega+v_F k),
    \eearr
and
  \bearr
      \xi =&&\frac{\omega^2+2 \omega v_F k -v_F^2 q^2-2 v_F^2 kq \cos \phi}{2(\omega+v_F k)}.
  \eearr
   In oder to do angular integration we consider $\xi$ as a function of $\phi$ and use $\delta[\xi(\phi) ]=\frac{\delta(\phi-\phi_0)}{|\partial_\phi \xi(\phi)|_{\phi_0}|}$ where $\cos\phi_0=-\vartheta$ then do integration on momentum $k$:
    \be
    \int d\phi \Im[P(\bsk,\bsq,\omega)]=\pi \frac{\gamma_{\vartheta}(k,\bsq,\omega)}{v_F^2 kq\sqrt{1-\vartheta^2}},
     \ee
     with
   \be
   \gamma_{\vartheta}(k,\bsq,\omega)=\omega+2  v_F k \sin^2 \theta +2v_Fk \vartheta^2\cos 2\theta- v_Fq \vartheta\cos 2\theta.
   \ee
   Here, the significant is limitation on the value of $\cos$ which causes 
   $-1\leq\vartheta\leq1$. Applying this constriction to  integration on momentum $k$ leads to the answer zero in the area with $\omega>v_f q$. By applying the same process to second term of $\Delta \Pi$ we get,
  \be
     \int d\phi \Im[P(\bsk,\bsq,-\omega)]=-\pi  \frac{ \gamma_{\vartheta^{\prime}}(k,\bsq,-\omega)sgn(\omega-v_F k)}{v_F^2 kq\sqrt{1-\vartheta^{\prime2}}},
      \ee
     where 
    \be
\vartheta'=\frac{v_F^2 q^2 -\omega^2+2 \omega v_F k}{2v_F^2 kq},
    \ee
and $\gamma_{\vartheta'}$ is otbained from $\gamma_\vartheta$ by substituting $\vartheta'$ for $\vartheta$. 
 Then the integration on $k$ space gives the following results:
\bearr
  &&\mp\frac{g_s g_v}{16 \pi \hbar} \Theta(|\omega|-v_Fq) 
    \left[C_1G\left( x\right)+C_2 \sin^{-1}\left( x\right)\right]^X_{X'}\nn\\
    &&\equiv  \left[G_1^{\pm\omega}(x)\right]^X_{X'},
\eearr
and
\bearr
     &&\pm\frac{g_s g_v }{16 \pi \hbar} \Theta(v_F q-|\omega|)
      \left[
     -C_1 G\left( x \right) + C_2 \cosh\left( x\right)\right]^X_0\nn\\
    &&\equiv \left[G_2^{\pm \omega }(x)\right]^X_0,
\eearr
where the definition of $C_1,C_2$ and $G, \cosh,\sin^{-1}$ are already given in Eqns.~\eqref{c1.eqn} to \eqref{sin-1.eqn}. 
The integration limits  $X$ and $X'$ will be determine by the value of $v_Fq$, $v_F k_F$ and  $\omega$.
Here the superscript $\pm \omega$ indicates whether the first and second term in Eq.\ref{app-deltapi}
is being dealt with.

\section{ Response of quantum matter to the transverse magnetic filed}
In this appendix we present details of the electromagnetic response of single layer of graphene to 
electromagnetic fields with emphasize on the tensor nature of the conductivity tensor. 
The following derivation holds for any quantum material which is specified by a two dimensional 
conductivity tensor $\sigma(\bsq,\omega)$. 
We assume that graphene is placed in $xy$ plane, and therefor the electric field in $z$ direction should be decaying 
away from the graphene plane~\cite{stern} i.e., $\bs{E}(\bsq,\omega)= \bs{E} e^{(i\bsq .\bs{r}-\zeta |z| -i\omega t)}$ with $\zeta=\sqrt{q^2-\omega^2/c^2}$.  In what follow, we proceed from Maxwell equations in Gaussian units for the rest of calculation. Let us start with,
\bearr
 \label{tm1}
 \nabla (\nabla.\vec{\bs{E}})-\nabla^2 \bs{E}=\frac{1}{c^2} \big(4\pi i  \omega \bs{\sigma}(\bsq,\omega)+ \omega^2\big) \bs{E}.
 \eearr
In the case of graphene conductivity is a tensor rather than a scalar which is represent by Eq.~\ref{current tensor}.
 Symmetry of Hamiltonian implies that $\sigma^{xy}(\bsq,\omega)=\sigma^{yx}(\bsq,\omega)$ and is defined by\cite{Dressel}
 \be
 \bs{J}(\bsq,\omega)=  \bs{\sigma}(\bsq,\omega) \bs{E},
 \ee
 with
 \be
  \label{current tensor}
 \bs{\sigma}(\bsq,\omega)= 
  \begin{pmatrix} \sigma^{xx}(\bsq,\omega) & \sigma^{xy}(\bsq,\omega)\\ \sigma^{yx}(\bsq,\omega) & \sigma^{yy}(\bsq,\omega) \end{pmatrix}
  \ee
Let us look at the Cartesian components of Eq.~\ref{tm1}. 
Along $x$ we have,
\bearr
&& -q_xq_y E_y+q_y^2 E_x+i q_x \frac{\pr E_z }{\pr z} -\frac{\pr^2 E_x}{\pr z^2}=\\&&\frac{1}{c^2}  \bigg[4\pi i  \omega \bigg( \sigma^{xx}(\bsq,\omega) E_x +\sigma^{xy}(\bsq,\omega) E_y\bigg)+ \omega^2 E_x\bigg]\nn.
 \eearr
Along $y$ it gives,
 \bearr
&&-q_xq_y E_x+q_x^2 E_y+i q_y \frac{\pr E_z }{\pr z} -\frac{\pr^2 E_y}{\pr z^2}=\\&&\frac{1}{c^2} \bigg[4\pi i  \omega \bigg( \sigma^{xy}(\bsq,\omega) E_x +\sigma^{yy}(\bsq,\omega) E_y\bigg)+ \omega^2 E_y\bigg],\nn
 \eearr 
 while along $z$ we have,
 \bearr
 \label{ezq}
&&i q_x\frac{\pr E_x }{\pr z}+ i q_y\frac{\pr E_y }{\pr z}+(q_x^2 +q_y^2)E_z=\bigg( \frac{\omega^2}{c^2} \bigg)   E_z.
 \eearr
Next using the plane wave form and the exponential decay in $z$ direction we substitute for the
partial derivatives to obtain
    \bearr
     \label{xtn}
     && -q_xq_y E_y-i q_x \zeta E_z +q_y^2 E_x -\zeta^2E_x=\\&&
     \frac{1}{c^2}  \bigg[4\pi i  \omega \bigg( \sigma^{xx}(\bsq,\omega) E_x +\sigma^{xy}(\bsq,\omega) E_y\bigg)+ \omega^2 E_x\bigg],\nn\\
     \label{ytn}
     && -q_xq_y E_x-i q_y \zeta E_z +q_x^2 E_y -\zeta^2E_y=\\
     &&\frac{1}{c^2} \bigg[4\pi i  \omega \bigg( \sigma^{xy}(\bsq,\omega) E_x +\sigma^{yy}(\bsq,\omega) E_y\bigg)+ \omega^2 E_y\bigg],\nn\\
     \label{ezqn}
       &&-i (q_x E_x+q_y E_y)+\zeta E_z=0.
       \eearr 
Combining Eq.~\ref{xtn} and Eq.~\ref{ytn} and inserting the value of $\zeta$ gives,
       \bearr
       \label{eyx1}
      \frac{ E_y}{ E_x}=  \frac{ q_y \sigma^{xx}(\bsq,\omega) - q_x \sigma^{xy}(\bsq,\omega)   }{ q_x \sigma^{yy}(\bsq,\omega)- q_y\sigma^{xy}(\bsq,\omega) }. 
          \eearr 
The continuity equation on the other hand gives,
 \bearr
 \int \bs{E}(\bsq,\omega).d\bs{S}=4 \pi \int \rho (\bsq,\omega) dV,
 \eearr
 or equivalently,
  \bearr
 E_z |_{z=0^+}- E_z |_{z=0^-}=4 \pi \int _{0^-}^{0^+} \rho (\bsq,\omega) dV. 
  \eearr
which then becomes,  
 \bearr
  \label{be1}
        E_z=\frac{2 \pi}{\omega}  (q_x J_x+q_yJ_y).
  \eearr
Combining this equation with Eq.~\eqref{ezqn} gives, 
    \bearr
    \label{eyx2}
        \frac{E_y}{E_x}= -\frac{ i q_x-\frac{2\pi \zeta}{\omega}( \sigma^{xx}(\bsq,\omega) + \sigma^{xy}(\bsq,\omega) )  }{ iq_y-\frac{2\pi \zeta}{\omega} ( \sigma^{xy}(\bsq,\omega)+q_y\sigma^{yy}(\bsq,\omega)) }.
           \eearr 
Now the comparison of this equation with~\eqref{eyx1} gives, 
    \bearr
    \label{tmf}
   1+\frac{2\pi i \sqrt{q^2-\omega^2/c^2}}{\omega} \varGamma'(\bsq,\omega)=0,
    \eearr
 where   
   \be
  \varGamma'(\bsq,\omega)=\frac{ \sigma^{xx}(\bsq,\omega)\sigma^{yy}(\bsq,\omega)-\sigma^{xy}(\bsq,\omega) \sigma^{xy}(\bsq,\omega)}{\hat{q}_{x}^{2} \sigma^{yy}(\bsq,\omega)+\hat{q}_y^2 \sigma^{xx}(\bsq,\omega)-2 \hat{q}_x\hat{q}_y \sigma^{xy}(\bsq,\omega)}.
   \ee
When the tensor character of conductivity tensor $\sigma$ is not important, i.e., when $\sigma^{xy}=\sigma^{yx}=0$ and
$\sigma^{xx}=\sigma^{yy}$ then the above equation reduces to the one used Ref.~\cite{mikhailove}.
Therefore the present equation properly encodes the tensor character of $\sigma$ into the propagation of 
electromagnetic waves in a quantum material.

 \section{ Response of quantum matter to transverse electric field}
 Response of a two dimensional quantum material whose quantum nature is encoded in the
 conductivity tensor $\sigma(\bsq,\omega)$ to transverse electric mode are similar to the transverse 
 magnetic case. 
 In what follow, we use  tensor representation Eq.~\ref{current tensor} and the Fourier transformation $\bs{H}(\bsq,\omega)=\bs{H} e ^{(i\bs{q}.\bs{r}-\zeta |z|-i\omega t)}$ plugged into the TE mode equation,
  \bearr
  \label{te}
  \nabla (\nabla.\bs{H})-\nabla^2 \bs{H}=\frac{1}{c^2} \big(4\pi i \omega \bs{\sigma}(\bsq,\omega)+ \omega^2 \big) \bs{H}
  \eearr
Rewriting the $x,y,z$ components of this equation gives,
 \bearr
 && -q_xq_y H_y-i q_x \zeta  H_z +q_y^2 H_x -\zeta ^2H_x=\nn\\&&
  \frac{4\pi i  \omega }{c^2} \bigg(\sigma^{xx}(\bsq,\omega) H_x+\sigma^{xy}(\bsq,\omega) H_y \bigg) +\frac{ \omega^2}{c^2} H_x,\\
 &&-q_xq_y H_x-i q_y \zeta H_z+q_x^2 H_y -\zeta^2 H_y=\nn\\
 &&\frac{4\pi i  \omega }{c^2} \bigg(\sigma^{xy}(\bsq,\omega) H_x+\sigma^{yy}(\bsq,\omega) H_y \bigg) +\frac{ \omega^2}{c^2} H_y,\\
 \label{etzq}
  && i( q_x H_x+ i q_y H_y)=\zeta  H_z.
 \eearr
The first two equations along  $x$ and $y$ directions give the ratio of $H_y$ to $H_x$,
  \bearr
  \label{h1}
      \frac{ H_y}{ H_x}=  \frac{ q_y \sigma^{xx}(\bsq,\omega) - q_x \sigma^{xy}(\bsq,\omega)   }{ q_x \sigma^{yy}(\bsq,\omega)- q_y\sigma^{xy}(\bsq,\omega) }. 
          \eearr 
Now starting from the Ampere's law we have,
  \bearr
 && \int \bs{H}.d\bs{L}=\frac{4 \pi}{c} \bs{J}_n(\bsq,\omega),  \\
  &&H_t|_{z=0^+} -H_t|_{z=0^-}=\frac{4 \pi}{c} \bs{J}_n(\bsq,\omega) ,
  \eearr
which give,
 \bearr
 \label{bh1}
       &&H_x=\frac{2 \pi}{c} \bigg(\sigma^{xy}(\bsq,\omega) E_x +\sigma^{yy}(\bsq,\omega) E_y\bigg )  ,\\
  \label{bh2}
       &&H_y=-\frac{2 \pi}{c}  \bigg(\sigma^{xx}(\bsq,\omega) E_x +\sigma^{xy}(\bsq,\omega) E_y\bigg )   . 
  \eearr
Using these two equations to construct the ratio $H_y/H_x$ and comparing it with  
Eq.~\eqref{h1} one must have 
\be
\frac{E_x}{E_y}=-\frac{q_y}{q_x}.
\label{rationExEy.eqn}
\ee

On the other hand for the $H_z$ we have
  \be
  H_z=\frac{c}{i\omega} \nabla\times 
  E|_z=\frac{c}{\omega}(q_x E_y-q_y E_x).
  \ee
Elimination of $H_z$ between this equation and Eq.~\eqref{etzq}, 
and substitution for the ratios of $E_x/E_y$ from Eq.~\eqref{rationExEy.eqn} and Eq.~\eqref{h1} gives,
 \bearr
 1-\frac{2\pi i \omega}{c^2 \sqrt{q^2-\omega^2/c^2}} \varGamma(\bsq,\omega) =0,
 \eearr
where  
 \bearr
 \varGamma(\bsq,\omega) =&&\hat{q}_x^2 \sigma^{yy}(\bsq,\omega)+\hat{q}_y^2 \sigma^{xx}(\bsq,\omega)\nn\\&& -2\hat{q}_x \hat{q}_y \sigma^{xy}(\bsq,\omega).
 \eearr
 
\section{damping structure}
Based on dispersion relation of TE and TM modes in Eq.~\eqref{te-simp} and  Eq.~\eqref{tm-simp}, 
the presence of imaginary part of $\Pi^{xx}(\bsq, \omega)$ causes a damped structure which can 
be lead to damping of the modes specified with a non-zero phase $\theta$ of the wave vector 
$q^\prime +iq^\dprime=q e^{i \theta}$ where $\tan\theta=\frac{q^\dprime}{q^\prime}$. 
Let us assume that the imaginary part (proportional to $\theta$) is small and expand
the TE and TM mode equations,
\begin{align}
   &1+\frac{2\pi e^2 }{c^2 \sqrt{(qe^{i\theta})^2-\omega^2/c^2}} \Pi^{xx}(q\hat{\boldsymbol{y}},\omega)=0,~~~~~\mbox{(TE)}\\
   &1-\frac{2\pi e^2 \sqrt{(qe^{i\theta})^2-\omega^2/c^2}}{\omega^2} \Pi^{xx}(q\hat{\boldsymbol{x}},\omega)=0,~~~~~\mbox{(TM)}
\end{align}
For the TE mode we have
\be
   c^2 \sqrt{(qe^{i\theta})^2-\omega^2/c^2}=-2\pi e^2 \Pi^{xx}(q\hat{\boldsymbol{y}},\omega)
\ee
which to first order in $\theta$ give,
\bearr
   &&c^2 \sqrt{q^2-\omega^2/c^2}=-2\pi e^2 \Re[\Pi^{xx}(q\hat{\boldsymbol{y}},\omega)],\\
   &&c^2 \frac{q^2\theta}{\sqrt{q^2-\omega^2/c^2}}=-2\pi e^2 \Im[\Pi^{xx}(q\hat{\boldsymbol{y}},\omega)].
\eearr
Combining the above equations we get,
\be
   q^\dprime q^\prime \simeq q^2\theta=(\frac{2\pi e^2}{c^2})^2 \Im[\Pi^{xx}(q\hat{\boldsymbol{y}},\omega)] \Re[\Pi^{xx}(q\hat{\boldsymbol{y}},\omega)].
\ee 
If we use dimensionless quantity of $Q$ and $\Omega$ and introduce  $\Pi^\prime(Q\hat{\boldsymbol{y}},\Omega) =\frac{16 \hbar}{e^2 g_s g_v} \frac{\Re[\Pi^{xx}(Q\hat{\boldsymbol{y}},\Omega)]}{v_F k_F} $ and $\Pi^\dprime(Q\hat{\boldsymbol{y}},\Omega) =\frac{16 \hbar}{e^2 g_s g_v} \frac{\Im[\Pi^{xx}(Q\hat{\boldsymbol{y}},\Omega)]}{v_F k_F}$ the damping structure of TE mode will be characterized by a small parameter $\theta$ that satisfies,
\be
(\frac{c}{v_F})^2  Q^\dprime Q^\prime \simeq (\frac{c}{v_F})^2 Q^2 \theta=(\frac{\pi \alpha}{2})^2 \Pi^\dprime(Q\hat{\boldsymbol{y}},\Omega) \Pi^\prime (Q\hat{\boldsymbol{y}},\Omega)
\ee 
where $\alpha$ is fine structure constant which is equal to $1/137$ and $v_F/c$ is much less than one which then
manage to give a very small imaginary part $Q^\dprime$ for the TE mode. 
The same result in dimensionless format is,
\be
   Q^\dprime=- \pi \frac{\alpha v_F }{2 c} \Im \Pi^{xx}(Q\hat{\boldsymbol{y}},\Omega).
\ee

We can repeat above process for TM mode as:
\begin{align}
\frac{\omega^2}{ \sqrt{(qe^{i\theta})^2-\omega^2/c^2}}=2\pi e^2  \Pi^{xx}(q\hat{\boldsymbol{x}},\omega)
\end{align}
which assuming that $\theta$ is small, to leading order gives,
\bearr
&&\frac{\omega^2}{\sqrt{q^2-\omega^2/c^2}}=2\pi e^2 \Re[\Pi^{xx}(q\hat{\boldsymbol{x}},\omega)],\\
&&\frac{ \omega^2 q^2 \theta }{(q^2-\omega^2/c^2)^{3/2}}=-2\pi e^2 \Im[\Pi^{xx}(q\hat{\boldsymbol{x}},\omega)].
\eearr
On the other hand for TM mode, the energies for a given wave vector are much smaller
than the energy of a photon in free space at the same wave vector. So for such a mode
we can ignore $\omega^2/c^2$ compared to $q$ which then give,
\be
   q^\dprime=-\frac{2\pi e^2 q^2}{\omega^2}\Im[\Pi^{xx}(q\hat{\boldsymbol{x}},\omega)].
\ee
which could be represented in terms of of dimensionless variable $Q$ and $\Omega$ as:
\be
   Q^\dprime=- \pi \frac{\alpha c }{2 v_F} \frac{Q^2}{\Omega^2} \Im\Pi^{xx}(Q\hat{\boldsymbol{x}},\Omega).
   \label{tmbroad.eqn}
\ee
Typically the ratio $c/v_F$ is on the scale of $10^2$ which cancels with the fine structure
constant $\alpha=1/137$ and therefore for small values of $\Im\Pi$ we end of for small broadening
which has been represented in Fig.~\ref{tm.fig}.

\section*{References}

\bibliography{mybibfile}

\end{document}